\documentclass[aps,prd,reprint,longbibliography, amsmath, amssymb, nofootinbib]{revtex4-2}
\usepackage[colorlinks=true, allcolors=black]{hyperref}

\usepackage[normalem]{ulem}
\usepackage{comment}
\usepackage[super]{nth}
\usepackage{graphicx}
\usepackage{dcolumn}
\usepackage{bm}
\usepackage[dvipsnames]{xcolor}
\usepackage{pgfplotstable, booktabs}
\usepackage{array}

\usepackage{amsmath}
\usepackage{xcolor}

\usepackage{float}
\usepackage{enumitem}

\renewcommand\apj{{ApJ}}
\newcommand\apjl{{ApJL}}     
\newcommand\aap{{A\&A}}
\newcommand\mnras{{MNRAS}}
\newcommand\pasp{{PASP}}

\UseRawInputEncoding

\begin{document}

\title{Effects of eccentricity on accreting binary black holes: MHD simulations in full GR reveal novel periodicities in jet power and synchrotron spectra}

\author{Vikram Manikantan}
\email{vik@arizona.edu}
\affiliation{Steward Observatory \& Department of Astronomy, University of Arizona, Tucson, AZ 85721, USA}

\author{Vasileios Paschalidis}
\email{vpaschal@arizona.edu}
\affiliation{Steward Observatory \& Department of Astronomy, University of Arizona, Tucson, AZ 85721, USA}
\affiliation{Department of Physics, University of Arizona, Tucson, AZ 85721, USA}

\author{Gabriele Bozzola}
\affiliation{Steward Observatory \& Department of Astronomy, University of Arizona, Tucson, AZ 85721, USA}
\affiliation{Division of Geological and Planetary Sciences, California Institute of Technology, Pasadena, 91125 California, USA}

\begin{abstract}
   We perform simulations of magnetohydrodynamic accretion onto equal-mass, non-spinning binary black holes in 3+1 full general relativity addressing the effects of orbital eccentricity. We find that binary black holes with non-negligible eccentricity accrete matter with periodicity that matches the binary orbital period, whereas quasicircular binaries exhibit accretion rate modulation at approximately $\sim 0.7\times$ their binary orbital period. Additionally, we find that the total jet luminosity is modulated at the orbital period for eccentric binaries, while quasicircular binaries only exhibit long term modulations. We perform a radiative transfer calculation of the dual jet synchrotron emission and demonstrate that the optically thin synchrotron emission varies on the binary orbital period for eccentric binaries. Moreover, eccentric binaries spend more time in a {\it low} state, where the synchrotron emission is minimum, than in a {\it high} state, where the synchrotron emission peaks. The quasicircular binary also exhibits variability in its optically thin synchrotron emission but the exact frequency of variability does not appear robust against different parameters. Our suite of simulations is an essential step towards providing a comprehensive catalog of multimessenger theoretical models that will enable studies of supermassive binary black holes detectable across the electromagnetic and gravitational wave spectra.
\end{abstract}

\maketitle

\section{Introduction} \label{sec:intro}
     Supermassive binary black holes (SMBBHs) are expected to form ubiquitously as the product of galaxy mergers~\cite{begelman_massive_1980, white_galaxy_1991,rodriguez_2009}. During the merger of the host galaxies, dynamical friction and three-body interactions drive the inward migration of the SMBHs down to parsec separations where they encounter the ``final parsec problem" -- where theoretical models find that MBBHs stall during their coalescence~\cite{begelman_massive_1980, volonteri_assembly_2003, milosavljevic_long-term_2003, milosavljevic_final_2003}. Recent studies have shown that this ``problem" can be solved by triaxiality in galaxy models or with self-interacting dark matter~\cite[see, e.g.,][]{gualandris_collisionless_2017, vasiliev_final-parsec_2015, alonso-alvarez_self-interacting_2024}. Once the binary orbital separation, $d$, is within a few milliparsecs, the binary inward migration is driven by gravitational waves, and the inspiral timescale becomes shorter than a Hubble time, $t_{\rm gw}\sim  1.45\times 10^{9} (d/5 {\rm mpc} )^4 (M/10^7 M_\odot)^{-3} \rm yrs$ for an equal-mass binary with total gravitational mass $M$~\cite{shapiro_black_1983}. When the binary orbital separation is within a few tens of gravitational radii ($d \sim 20-30 r_g$ where $r_g\equiv GM/c^2$ is the gravitational radius), the binary is in the strong-field dynamical spacetime regime and the inspiral timescale for equal-mass binaries is $t_{\rm gw}\sim 36 \, (d/30r_g)^4 (M/10^7 M_\odot) \rm \, days$. This is the regime that gravitational wave detectors target, and is the focus of our work.

    The inspiral and merger of at least a fraction of SMBBHs is expected to take place in gaseous environments. The gas-rich post-galaxy merger environment leads to an increased rate of tidal disruption events, star formation, and SMBH activity~\cite{rodriguez_2009, chen_enhanced_2009, li_direct_2019}. Such a gas rich environment makes SMBBHs excellent candidates for multi-messenger astronomy because: i) gas accretion onto the SMBBH will drive emission across the electromagnetic (EM) spectrum (see~\cite{bogdanovic_bhb_review} for a recent review), and ii) the inspiral of the SMBBH generates detectable gravitational waves (GW)~\cite{LISA_white_paper, lisa_gw_whitepaper, amaro-seoane_astrophysics_2023}.

    However, observing SMBBHs is challenging. Direct GW observations of SMBBHs are not currently possible because there exist no GW detectors that are sensitive enough to probe gravitational waves from individual SMBBHs in the $n\rm Hz$ to $\rm Hz$ bands\footnote{For quasicircular binaries the GW frequency is $f_{\rm GW} \sim 7.2 \times 10^{-5} (M/10^7 M_{\odot})^{-1} (d/20r_g)^{-3/2} \, \rm Hz$}. The Laser Interferometer Gravitational-Wave Observatory and  Virgo detectors have peak sensitivity at $\sim 100 \, \rm Hz$ and thereby observe stellar-mass black hole mergers~\cite{gw150914}. On the other hand, Pulsar Timing Arrays (PTA) observe in the $n\rm Hz$ regime, which is appropriate for $10^{9-10} M_{\odot}$ binaries, but they have so far only yielded strong evidence for the existence of a stochastic gravitational wave background~\cite{nanograv}. Some studies have suggested that GWs by individual SMBBHs could be detectable by current generation PTAs, however, the expected 20-year timeline of such a detection is longer than the current 15-year dataset available to PTAs~\cite{rosado_expected_2015, kelley_single_2018}. Furthermore, previous PTA observations placed limits on the detection of individual sources: no observable $10^9 M_\odot$ binaries within 120 Mpc and no $10^{10} M_\odot$ binaries within 5.5 Gpc at their most sensitive sky location~\cite{aggarwal_nanograv_2019}. Adding more high precision pulsars in the future will increase the PTA sensitivity $\propto \sqrt{N_{\rm pulsars}}$, and hence increase the chances of detecting individual SMBBHs (see, e.g., Fig.~4 in~\cite{babak_forecasting_2024}). However, we may have to wait until the mid 2030s for the launch of the space-based Laser Interferometer Space Antenna (LISA) to observe in the mHz to Hz GW regime where we expect loud SMBBHs~\cite{LISA_white_paper, lisa_gw_whitepaper, arun_new_2022, amaro-seoane_astrophysics_2023}. 
    
    On the other hand, EM surveys have discovered more than 200 SMBBH candidates at varying orbital separations~\cite{rodriguez_2006, rodriguez_2009, charisi_multiple_2015, graham_systematic_2015, panstars_smbh, oneill_unanticipated_2022, kiehlmann_pks_2024}. These candidates include a spatially resolved close-separation SMBBH ($7.3 \, \rm pc$) with very long baseline interferometry~\cite{rodriguez_2006}, radio observations of jet rotation in OJ287 that could be attributed to a secondary BH~\cite{radio_oj287_jet}, Doppler shifted broad line emission possibly from the orbital motion of a SMBBH~\cite{runnoe_radial}, a TDE that can be explained by a SMBBH~\cite{wen_at2018fyk_2024}, and directly observing a dual AGN with separations greater than a kiloparsec~\cite{komossa_discovery_2003, comerford_dualagn}. However, the vast majority of these candidates are not in the strong-field dynamical spacetime regime.
    
    There are also more than 25 SMBBH candidates with a total mass of $ > 10^9 \, M_{\odot}$ and orbital separation $\lesssim 2-4 \, \rm mpc$~\cite{panstars_smbh}. This puts them in the GW-driven and in the strong-field dynamical spacetime regime (see Fig.~1 in~\cite{bright_minidisk}). Furthermore, future observatories, such as the Roman Space Telescope and Athena, are poised to uncover several hundred SMBBHs candidates out to redshift $z=6$ with masses in the observing capabilities of LISA~\cite{haiman_massive_2023, piro_chasing_2023}. However, confirmed observations of SMBBHs at the sub-parsec scale remain elusive; we would need confirmation by a GW detection, a direct observation of the SMBBH event horizons, or a ``smoking gun" EM signature of a SMBBH. To aid with this continued search, the community has turned to theoretical modeling and direct numerical simulations to identify EM signals that are unique to SMBBHs. This work focuses on this exact goal.
    
    SMBBHs are likely surrounded by hot magnetized plasma. If the gas around the SMBBH has enough angular momentum, it will circularize to form a circumbinary disk (CBD)~\cite{barnes_formation_2002}. The SMBBH resides at the center of this CBD in a lower density cavity that has been cleared out by the binary tidal torques~\cite{artymowicz_dynamics_1994, macfadyen_eccentric_2008}. Gas from the CBD then accretes onto the binary through two tidal streams that can potentially form minidisks around the individual black holes~\cite{gold_accretion_2014, farris_binary_2014, bowen_relativistic_2017,paschalidis_minidisk_2021, bright_minidisk}. These minidisks then accrete onto the BHs, generating emission in the x-ray and UV, producing doppler shifted emission lines, and powering relativistic jets if the fluid is carrying a magnetic field to the BH horizons (see e.g.,~\cite{Palenzuela:2010nf,Moesta:2011bn, bright_minidisk,avara_accretion_2023}, and~\cite{bogdanovic_bhb_review} for a recent review on electromagnetic signatures of SMBBHs). 
    
    There are different computational approaches to studying these systems, each one applicable and valid in its respective regime. Newtonian 2D hydrodynamic studies can evolve binaries for thousands of orbits and help understand the widely-separated non-relativistic regime (see e.g.,~\cite{munoz_circumbinary_2020, westernacher-schneider_multi-band_2022, lai_circumbinary_2023, delaurentiis_relativistic_2024} and references therein for some recent work). Implementing post-Newtonian approximate background metrics allows for modeling of the SMBBH inspiral at closer separations than Newtonian approaches and can help gain some intuition into horizon-scale accretion~\cite{avara_accretion_2023, porter_parameter_2024}. However, performing reliable simulations from first principles and without approximations requires the solution of the Einstein equations, coupled to general relativistic magnetohydrodynamics, and radiation transport. Approaches that solve the Einstein equations are commonly referred to as fully general-relativistic (for reviews on fully general-relativistic work, see~\cite{gold_relativistic_2019,Cattorini:2023akr}).
    
    More recent studies of accretion onto BBHs in full general relativity have explored effects of the disk thickness~\cite{khan_disks_2018}, unequal mass ratios~\cite{gold_accretion_2014, ruiz_unequal_2023}, prograde and retrograde spins~\cite{paschalidis_minidisk_2021, bright_minidisk}, and misaligned spins~\cite{cattorini_misaligned_2022, ruiz_unequal_2023, fedrigo_grmhd_2024}, amongst other parameters. However, the impact of orbital eccentricity on SMBBH accretion in the relativistic regime has only been studied for a single simulation setup, in our previous work~\cite{Manikantan_BBH_eccentric_2024_letter}. Here, we expand on the details of our previous work and perform a comparison of CBD accretion onto quasicircular BBHs and BBHs with different values of eccentricity.
    
    While GWs are expected to radiate away orbital eccentricity~\cite{peters_matthews}, binary interactions with a CBD can maintain non-negligible eccentricity into the LISA band~\cite{gualandris_eccentricity_2022,valli_long-term_2024, franchini_behaviour_2024}. This eccentricity depends, amongst other parameters, on the binary-disk decoupling radius, which shrinks with increasing disk scale height, $H/r$~\cite{farris_binary_2012}. A smaller decoupling radius would allow the matter to torque the binary further into its evolution. For example, a thin disk ($H/r < 0.01$) has a decoupling radius $\mathcal{O}(200\,M)$, in which case the binary could radiate away its eccentricity before it enters the LISA band and before it inspirals to the separations that we are concerned with ($d < 30 \, M$). However, a thicker disk ($H/r\sim 0.1$, relevant for slim disks~\cite{SlimDisksAbramovich}) can have a decoupling radius as small as $30\, M$, allowing the binary to sustain a higher eccentricity~\cite{farris_binary_2012}. In fact, recent Newtonian hydrodynamic studies of BH binaries argue that eccentricity may be the norm for SMBBHs and can reach values up to $e\simeq 0.3-0.5$ in observable GW bands~\cite{roedig_limiting_2011,siwek_2024}. Thus, it is possible that at least a fraction of LISA binaries will have non-zero residual eccentricities in-band. Additionally, the presence of a third massive body, e.g., following a triple galaxy merger~\cite{yadav_2021,Ni_2022}, can result in binary eccentricity of order unity at relativistic separations (relevant for LISA) through chaotic, non-hierarchical three-body interactions~(see \cite{2018MNRAS.473.3410R} and references therein). 
    
    Furthermore, if the stochastic GW background comes from an eccentric population of MBBHs, future PTA experiments might detect eccentric MBBH merger events~\cite{eccentric_pta}. Moreover, eccentric binaries can probe more relativistic velocities than quasicircular binaries, thereby allowing to probe gravity in a regime not accessible by zero eccentricity. The detection of even one SMBBH binary at finite eccentricity could be a gold mine for understanding relativistic astrophysics and gravity. Therefore, establishing the effect of orbital eccentricity on accreting SMBBHs is very important.
    
    In this work we perform $3+1$ fully general-relativistic, magnetohydrodynamic simulations of relativistic BBHs on eccentric orbits with initial eccentricities of $e = {0.00, 0.17, 0.31}$, which are within the range predicted by~\cite{roedig_limiting_2011,siwek_2024}. We focus on answering the following key questions: i) how is the measured mass accretion rate and its periodicity affected by orbital eccentricity? ii) How is the measured jet luminosity and its synchrotron emission affected by orbital eccentricity? iii) Do eccentric binaries at relativistic separations form persistent minidisks? 
    
    This paper is structured as follows: in Section \ref{sec:approach} we outline our numerical methods and diagnostics, in Section \ref{sec:results} we describe our results, in Section \ref{sec:synchrotron} we perform an approximate radiative transfer calculation of synchrotron emission within the jet, in Section \ref{sec:coincident} we discuss simultaneous GW and EM emission from our models, and in Section \ref{sec:summary} we summarize our findings.
    
    Throughout, we adopt geometrized units in which $G = c = 1$, where $G$ is the gravitational constant and $c$ is the speed of light. Our spatial and time domains are measured in units of $M$, where $M$ is the Arnowitt-Deser-Misner (ADM) mass of the spacetime. In the spatial domain\footnote{$GM/c^2 = 1.48\times 10^{7} (M/10^7 M_\odot)\, \rm km$} $1 \, M = GM/c^2$ and in the time domain\footnote{$GM/c^3 = 49.25 \, (M/10^7 M_\odot) \, \rm seconds$} $1 \, M = GM/c^3$.


\section{Methods} \label{sec:approach}  

        \subsection{Initial data}
        In this section we discuss how we prepare the spacetime and matter initial data for our simulations.
                
        \subsubsection{Spacetime}

        We use the {\tt TwoPunctures} thorn to prepare initial data for the spacetime~\cite{ansorg_single-domain_2004,Paschalidis:2013oya}. We initialize our BHs with no spin and identical mass ($q = m_1/m_2 = 1$) at apoapsis. We choose a major axis $a/M = 20$ such that the initial orbital separation is determined by the initial eccentricity, $d = a(1+e)$, which we target to be $e=0, 0.15, 0.3$. We do this to ensure that the binaries have approximately the same initial orbital period, which we measure to be the case to within $10\%$. We introduce eccentricity into our BBH by first computing the 3rd order Post-Newtonian linear momenta corresponding to a quasicircular BBH and then adjusting their tangential component by a factor of $\sqrt{1-e}$, where $e$ is our target eccentricity. This approach would result in actual eccentricity equal to the target eccentricity if the binary was Newtonian. In our case this approach produces a value for the eccentricity that is within $\sim 3-10\%$ of the target value. Thus, once we set up initial data, we evolve the BBH before measuring our eccentricity (see Appendix \ref{sec:eccentricity_measurements} for details on how we measure the eccentricity). In this work we do not iteratively try to find the appropriate factor $\sqrt{1-e}$ which yields our exact target eccentricities, although this would be straightforward using a bisection approach. The measured eccentricities for the binaries evolved in this work are $0.00, 0.17, 0.31$. 

        \subsubsection{Matter and magnetic fields}
        We use the power-law torus solution for the initial conditions of our CBD as previously described in~\cite{gold_accretion_2014} and~\cite{khan_disks_2018}. The torus would be an equilibrium solution around a single BH with the same mass as the BBH gravitational mass. We set the inner edge of the CBD at $r/M = 18$ with specific angular momentum of $l = 5.15$. These parameters are used to set the outer edge of the disk at $r/M \simeq 100$. We use a $\Gamma$-law equation of state, with  $\Gamma = 4/3$, which is appropriate for radiation pressure dominated flows, even at $1/10$ of the Eddington accretion rate~\cite{liska_2T}. The initial vertical scale height of the circumbinary accretion disk is $H/r \simeq 0.24$, where $r$ is measured from the center of mass of the binary. We initialize our disk with a seed poloidal magnetic field as in~\cite{khan_disks_2018}, with minimum initial plasma beta $\beta = P_{\rm gas}/P_{\rm mag} \simeq 13$. While the initial minimum of the plasma beta chosen may be low compared to single black hole accretion studies, it is consistent with previous simulations in full general relativity. Moreover, 
        simulations with initial, $\beta \sim 100$, relax to $\beta \sim 10$ through the MRI~\cite{2013ApJ...767...30B}. Additionally, to resolve the MRI wavelength with our chosen resolution, we are limited to stronger initial magnetic fields. Despite our lower initial $\beta$, the initial net magnetic flux in the disk is only $\sim 3\times $ higher compared to an initial $\beta=100$, because the flux scales with $b$, while $\beta$ scales with $b^2$. Thus, the net vertical flux in the disk is not much larger than that  typically adopted in large scale simulations of accretion disks onto single black holes The initial maximum magnetization in our disk ($\sigma \equiv b^2/2\rho$) is $\sigma \simeq 4 \times 10^{-5}$, where $b$ is the magnitude of the magnetic field measured by an observer comoving with the fluid normalized by $\sqrt{4\pi}$. Therefore, even with our lower initial $\beta$ the  magnetic field is dynamically unimportant initially. Due to the seeded magnetic field, our disks are unstable to the magnetorotational instability (MRI)~\cite{Balbus}. We have checked that the wavelength the fastest growing MRI mode is smaller than the disk scale height, and that we resolve this wavelength with at least 20 zones in the bulk of the disk and a maximum of 50 zones at the inner edge of the disk.

     \subsection{Evolution} \label{sec:numerical}

        We employ the $3+1$ general relativistic, dynamical spacetime  magnetohydrodynamic (GRMHD), adaptive-mesh-refinement (AMR) \verb|IllinoisGRMHD| code~\cite{ilgrmhd} within the framework of the \verb|Einstein Toolkit|~\cite{einsteintoolkit}, which employs the Cactus/Carpet infrastructure for mesh refinement~\cite{Goodale2002a, schnetter_carpet_2016}. \verb|IllinoisGRMHD| evolves the matter and magnetic fields using the equations of ideal magnetohydrodynamics in flux-conservative form  via a high-resolution shock capturing scheme based on the HLL Riemann solver~\cite{toro_riemann_2009} and the Piecewise-Parabalic reconstruction~\cite{colella_piecewise_1984}. These methods have been described in~\cite{ilgrmhd}, and have  been extensively tested against other codes in~\cite{porth_event_2019}. For our EM gauge choice, we utilize the generalized Lorenz gauge condition from~\cite{farris_binary_2012} and set the Lorenz gauge damping parameter to $\xi=8/M$ to remove spurious gauge modes~\cite{Etienne:2011re}.
        
        We evolve the spacetime by solving Einstein's equations of general relativity in the Baumgarte-Shapiro-Shibata-Nakamura (BSSN) formalism~\cite{Shibata1995, Baumgarte1998}, as implemented in the \verb|LeanBSSN| thorn ~\cite{sperhake_binary_2007}.  We adopt the moving puncture gauge conditions~\cite{baker_gravitational-wave_2006, campanelli_accurate_2006} with the shift vector parameter $\eta$ set to $\eta=1.4/M$. 

        We use a three-dimensional cartesian grid with the outer boundary extending from $-5120 \, M$ to $+5120 \, M$ in the $x, y,$ and $z$ directions, with a total of 14 refinement levels. We adopt \verb|Carpet|~\cite{schnetter_carpet_2016} for adaptive mesh refinement (AMR), which implements a box-in-box AMR scheme. We use 3 sets of nested refinement levels, two of which are centered onto each puncture and one on the binary center of mass. The grid spacing on the coarsest (finest) refinement level is  $\Delta x = 128 \, M$ ($\Delta x=M/64$).
        The adopted resolution was based on: 1. requirements to resolve the wavelength of the fastest growing MRI mode, and 2. through a convergence study of the evolution of vacuum binaries, ensuring that, at the chosen resolution, the orbital separation and phase deviate by no more than $\mathcal{O}(0.1\%)$ from our highest-resolution runs over the first four orbits.

        We do not treat radiative feedback, heating, or cooling. Finally, the fluid does not backreact onto the spacetime, since the spacetime mass/energy content is dominated by the SMBBH.

        We evolved the $e = 0.00$ and $e=0.17$ binaries out to $t/M \sim 10,000$ (19 and 18 orbits, respectively) and the $e=0.31$ binary to $t/M \sim 6500$ (13 orbits). The $e=0.31$ binary inspirals significantly faster than our other cases, which changes the orbital frequency rapidly. Since the focus of this work is the quasistationary phase of the accretion, we stop the $e=0.31$ simulation earlier.

    \subsection{Diagnostics} \label{sec:diagnostics}
        We adopt the same diagnostic tools as in~\cite{bright_minidisk} to measure the rest-mass accretion rate ($\dot{M}$), Hill sphere rest-masses ($M_{\rm Hill}$), and outgoing Poynting flux. We locate apparent horizons with \texttt{AHFinderDirect}~\cite{Thornburg2004}. We use the package \verb|kuibit|~\cite{kuibit} for all our analysis.
        
        \subsubsection{Fourier Analysis} \label{subsubsec:Fourieranalysis}
            We perform Fourier analysis on time series, such as the BBH rest-mass accretion rate, and jet luminosity. Here we outline the steps of that Fourier analysis:

            \begin{enumerate}[noitemsep]
                \item \textit{Obtain} the time series output from the simulation.
                \item \textit{Remove} the running average of the time series to remove any DC components.
                \item \textit{Window} our signals using a Hann window.
                \item \textit{Zero-pad} our signal such that the total number of points in the signal is $2^{18}$. This choice is arbitrary, but ensures an even number of data points in the signal.
                \item \textit{Perform a fast Fourier transform} on the signal using the \verb|scipy.fft| function~\cite{scipy}.
                \item \textit{Normalize} the power spectral density (PSD) to a maximum value of one and re-scale the frequency axis by the orbital frequency. 
            \end{enumerate}

        \subsubsection{Orbital Frequency} \label{subsubsec:measuringfrequency}
             We normalize all Fourier frequencies to the orbital frequency of our BBHs. To compute the orbital frequency in a gauge-invariant way, we use the GW signals generated by our binaries, which we extract using the \texttt{NPScalars} thorn of the Canuda suite~\cite{canuda,canudacode}. In the quasicircular case, the BBH orbital frequency equals $1/2$ the frequency of the $\ell=2,m=2$ gravitational wave mode. We compute this from the first time derivative of the unfolded phase of the $\ell=2,m=2$ gravitational wave strain, $\varphi_{22}$, where the orbital frequency is given by $f_{22}\equiv\frac{d\varphi_{22}}{dt}$.  This frequency remains approximately constant in time in our quasicircular  case. However, the GW frequency for eccentric orbits has time dependence even on an orbital time (higher frequency near pericenter and lower frequency at apocenter). Therefore, to recover the orbital frequency in the eccentric cases -- defined as the time between consecutive pericenter passages -- we perform a Fourier transform of the $f_{22}$ time series. This frequency oscillates on the orbital timescale and its Fourier transform reveals the orbital frequency of the binary. The orbital period of our eccentric binaries evolves rapidly due to GW emission, especially in our high eccentricity case ($e=0.31$) (see Appendix \ref{sec:eccentricity_measurements}). Therefore, when we normalize Fourier frequencies by the binary orbital frequency, we specify the time at which we perform the Fourier transform and measure the orbital period, which is the same time period unless otherwise stated. 
            


        \subsubsection{Hill Sphere} \label{subsubsec:hill}
            The Hill sphere or sphere of gravitational influence around a body is given by~\cite{hamilton_orbital_1991}

            \begin{equation}
                r_{\rm Hill} = \frac{r_p}{2}\biggr(\frac{m_1}{3(m_1 + m_2)} \biggr)^{1/3}
            \end{equation}
            where $r_p$ is the pericenter orbital separation, $m_1$ is the mass of the object whose Hill sphere we are determining, and $m_2$ is the mass of the binary companion. As a reminder, in our simulations $m_1 = m_2$. Notice that these Hill spheres are defined by the separation at pericenter, but the effective Hill sphere is larger at apocenter. We use the volume contained within a coordinate radius $r_{\rm Hill}$ to compute the rest-mass of the minidisks around each BH, using the \verb|VolumeIntegralsGRMHD| thorn~\cite{duez_relativistic_2005}. We note that the Hill sphere radius calculation is neither gauge invariant nor relativistic. These Hill sphere radii serve as a crude radius of influence for each BH and they are only used to provide some intuition when analyzing our simulations.
            
\section{Results} \label{sec:results}
    
    \begin{figure*}
        \includegraphics[width=\textwidth]{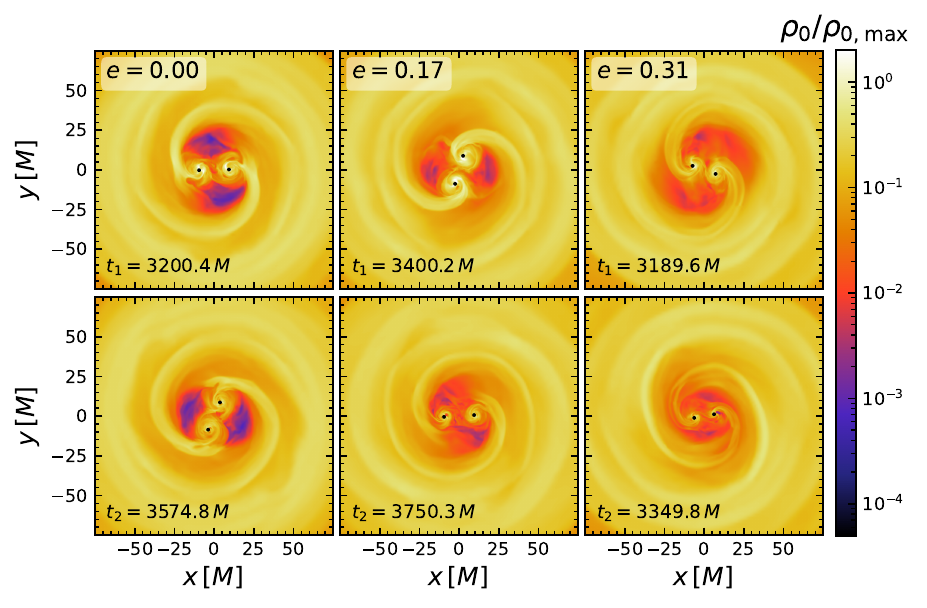}
        \caption{Rest-mass density ($\rho_0$) contours of each simulation at two snapshots: when the Hill sphere rest-mass is maximum ($t_1$, top row) and minimum ($t_2$, bottom row) (see maxima and minima of the green dashed line in Fig.~\ref{fig:mdot}). From left to right, we plot the eccentricities $e = 0.00, 0.17,$ and $0.31$. In each panel, we plot the gas rest-mass density normalized to $\rho_{0, \rm max}$, where $\rho_{0, \rm max}$ is the maximum density in the initial data and $\rho_{0, \rm max} = (1.6, 1.8, 2.6) \times 10^{-11} (\dot{M}/0.1 \dot{M}_{\rm edd}) (M/10^7 M_{\odot})^{-1} (\eta/0.1)^{-1} \rm \, g \, cm^{-3}$ for the $e=0.00, 0.17$ and $0.31$ binaries, respectively. The quasicircular binary ($e=0.00$, left column) harbors minidisks with higher density around each BH ($\rho_0/\rho_{0,{\rm max}} \sim 1$) which exist in a lower-density cavity ($\rho_0/\rho_{0,{\rm max}} \sim 10^{-3}$) than in the eccentric cases. This minidisk and cavity structure persists in time. The eccentric binaries ($e=0.17, 0.31$, center and right columns) also exist in cavities but the minimum density areas ($\rho/\rho_0 \sim 10^{-2}$) are less defined. Furthermore, they only have minidisks in the $t_1$ snapshot. In the $t_2$ snapshots, the minidisks are almost completely depleted. The black disks indicate the BH apparent horizons.}
        \label{fig:rho_panel}
    \end{figure*}
    
    In this section we describe the evolution of the MHD fluid, the accretion onto the BBH (Section \ref{subsec:accretion}), and the jet outflows (Section \ref{subsec:jets}). 
    
    \subsection{Accretion onto the Binary} \label{subsec:accretion}

        After initializing the circumbinary torus around our BBHs as described in Section \ref{sec:approach}, we evolve the systems until the accretion rate onto the BBH relaxes. As in previous BBH-CBD studies in full general relativity, we define relaxation (or quasisteady state) as when the amplitude and phase of the accretion rate are approximatelly constant~\cite{gold_accretion_2014,  paschalidis_minidisk_2021, bright_minidisk, ruiz_unequal_2023}. The BHs continue to reside in a lower-density cavity inside the CBD. The binary gravitationally torques the CBD and matter near the inner disk edge falls onto each BH through tidal streams (Fig.~\ref{fig:rho_panel}). The infalling gas briefly circularizes around each BH, forming a minidisk, before being accreted. This process repeats and modulates the mass accretion rate.  

        \begin{figure*}
            \includegraphics[width=\textwidth]{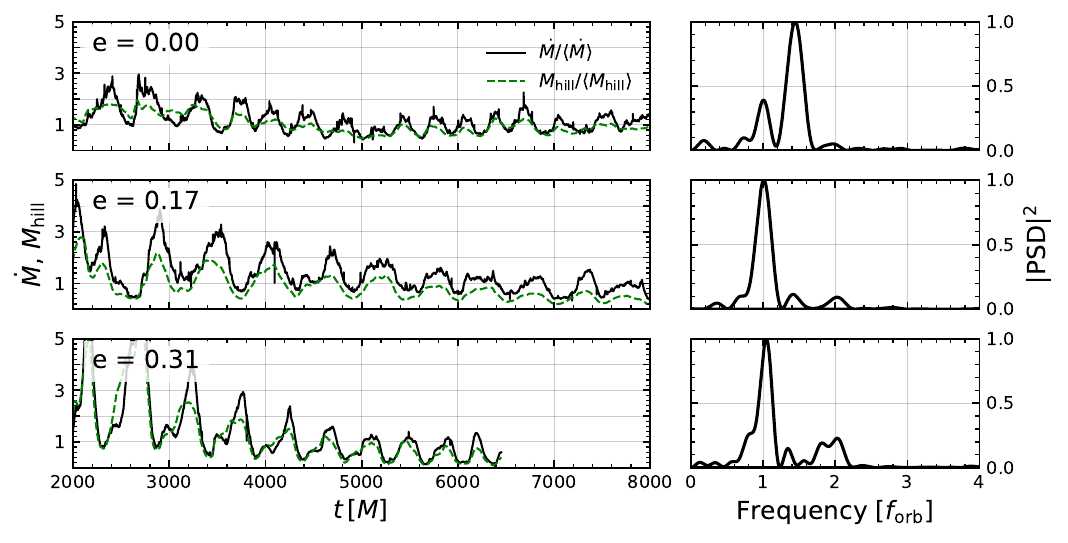}
            \caption{Left column: total rest-mass accretion rates (solid black line) onto both black holes and total rest-mass contained within the BH Hill spheres (dashed green line) vs time. Both quantities normalized by their averages for $t>3000 \, M$. Right column:  power spectral density (PSD) of the Fourier transform of the rest-mass accretion rate with the frequency normalized to the orbital frequency. For comparison, the Fourier transform is performed over the time period $t=3000-5500 \, M$ for all three binaries, limited by the rapid evolution of the $e=0.31$ binary orbit. The plots demonstrate that the dominant frequencies of the accretion are variability are $\sim f_{\rm orb}$ for eccentric binaries and $\sim 1.4\, f_{\rm orb}$ for quasicircular binaries. The dominant frequency is insensitive to our choice of over which time period we perform the Fourier transform, however, the secondary peak at $2 \, f_{\rm orb}$ in the eccentric cases vanishes after $5000 \, M$.}
            \label{fig:mdot}
        \end{figure*}

        In Fig.~\ref{fig:rho_panel}, we plot the gas rest-mass density on the $x-y$ plane (the orbital plane of the BBHs). We indicate the BH apparent horizons with black disks. The gas density is normalized to the initial maximum gas density in the torus, $\rho_{0,\rm max}$, which, for our $e=0.00$ binary, scales as\footnote{The $e=0.17$ and $e=0.31$ cases have the same scaling, but the numerical factor in front of the scaling relation as in Eq. \eqref{rho_scaling} is $1.8 \times 10^{-11}$ and $2.6 \times 10^{-11}$, respectively.}

        \begin{widetext}
            \begin{equation}\label{rho_scaling}
                        \rho_{0, \rm max} = 1.6 \times 10^{-11} \biggr(\frac{\langle\dot{M}\rangle}{0.1 \dot{M}_{\rm edd}}\biggr) 
                        \biggr(\frac{M}{10^7 M_{\odot}}\biggr)^{-1} 
                        \biggr(\frac{\eta}{0.1}\biggr)^{-1} \rm \, \frac{g}{ \, cm^{3}}.
            \end{equation}
        \end{widetext}
        Here, $\langle\dot{M}\rangle$ is the average total accretion rate after it settles ($t/M>3000$), $\dot{M}_{\rm edd}$ is the Eddington accretion rate, and $\eta$ is the radiative efficiency of the accretion flow. From left to right, we present the $e=0.00$, $e=0.17$, and $e=0.31$ cases, respectively. The top row shows moments in the simulation where the Hill spheres of the BHs are most filled ($t_{\rm 1}$) and the bottom row shows moments where the Hill spheres are depleted ($t_{\rm 2}$) (see the maxima and minima of the dashed green lines in Fig.~\ref{fig:mdot}).

        Figure \ref{fig:rho_panel} illustrates the quasiperiodic morphology of the eccentric systems. In the $e=0.00$ case, the $t_{\rm 1}$ and $t_{\rm 2}$ panels are not noticeably different. The approximately constant orbital separation of the quasicircular binary allows more persistent minidisk structures to form in the Hill sphere of each BH. Despite that, the minidisks undergo quasiperiodic oscillations in mass and accretion rate. Furthermore, the cavity carved by the BBH has a persistent low density area, with density $\rho_0/\rho_{0, \rm max} \sim 10^{-3}$. For quasicircular binaries, Newtonian 2D studies show that the cavity typically has a radius of $\sim 2a$, where $a$ is the BBH orbital separation. In full GR it has been shown that the cavity radius is $\sim a$ when no cooling is applied, and $\sim 1.5a$ with cooling~\cite{gold_accretion_2014}. This difference between 2D viscous studies and 3D MHD studies can be explained primarily by the overflow of the binary tidal barrier when the 3rd dimension appears. The thinner the disk is the stronger the tidal torque pushing matter away from the orbit becomes, see e.g.,~\cite{2009MNRAS.398.1392L}. In our eccentric cases the cavity radius is $\sim a$, but we plan to further investigate the cavity properties for eccentric binaries in a future work.
        
\begin{figure*}[ht]
            \includegraphics[width=\textwidth]{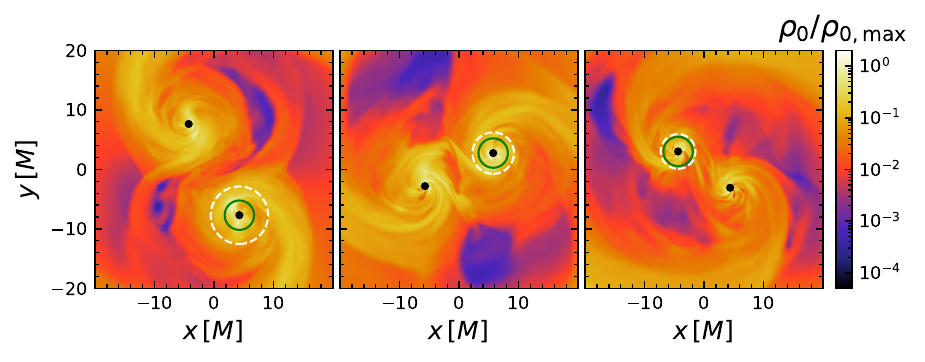}
            \caption{Contours of rest-mass density ($\rho_0$) for our $e=0.31$ simulation, normalized by the initial maximum density $\rho_{0, \rm max} =  2.6 \times 10^{-11} (\dot{M}/0.1 \dot{M}_{\rm edd}) (M/10^7 M_{\odot})^{-1} (\eta/0.1)^{-1} \rm \, g \, cm^{-3}$. The left panel corresponds to just after apocenter ($t/M=4120$), the middle panel as the binary approaches pericenter ($t/M=4220$), and the right at pericenter ($t/M=4300$). Around one of the BHs we indicate the innermost stable circular orbit (ISCO) with a solid green line and the Hill sphere radius with a dashed white line. Note that the ISCO coordinate radius is not a gauge invariant quantity. Here, it is set to its value in puncture coordinates (the gauge conditions we adopt) of a single non-spinning BH~\cite{Baumgarte:2007ht}: $r_{\rm ISCO}\simeq4.95m_1\simeq 2.475M$.
            We indicate the BHs horizons with black disks. The left panel shows that the accretion streams approaching the BHs form minidisks, which begin to be accreted in the middle panel, and are depleted on the right because the Hill sphere and ISCO radii almost coincide.}
            \label{fig:density_inset}
        \end{figure*}

        In the eccentric binaries, $e=0.17$ and $e=0.31$, the difference between the $t_{\rm 1}$ and $t_{\rm 2}$ panels in Fig!\ref{fig:rho_panel}  is pronounced. In the top row of the eccentric binaries ($e=0.17, 0.31$), there are minidisks of high density around each BH. However, in the bottom row, these minidisks are almost completely depleted. This can be explained by the change in the relative size of the Hill sphere with respect to the innermost stable circular orbit (ISCO) (Fig.~\ref{fig:density_inset}, more later).

        In Fig.~\ref{fig:mdot}, we analyze the rest-mass accretion rate. In the left column, we plot the time-dependent total rest-mass accretion rate onto the BHs with a black solid line and plot the rest-mass contained within the Hill sphere (as measured at pericenter) with a dashed green line. In the right panels, we plot the Fourier transform of the total rest-mass accretion rate for the time range $3000 < t/M < 5500$. We choose this time interval such that the accretion rate has reached a quasisteady state, but the interval is not so long that the rapid inspiral of the $e=0.31$ binary changes the  orbital frequency appreciably, so we can use it to normalize our periodograms. We fix this time interval for comparison across all of the simulations. The dominant frequency of the accretion rate variability for the quasicircular binary is $\sim 1.4 \times f_{\rm orb}$, and for the eccentric binaries is their orbital frequency, $f \sim f_{\rm orb}$. This is consistent with recent Newtonian hydrodynamic studies~\cite{westernacher-schneider_multi-band_2022, delaurentiis_relativistic_2024}. However, in those studies, where the separation of the BBH is $> 150 M$ and hence $r_{\rm Hill} \gg r_{\rm ISCO}$, it is possible that the matter inflow time from magnetized minidisks is much longer than the orbital period, which would quench the accretion rate variability~\cite{bright_minidisk}. Furthermore, we observe a secondary frequency at $2 f_{\rm orb}$ in the power spectrum of both eccentric binaries. We can see this feature in the timeseries by the double peaks in the early accretion rate $3000-5000\,M$ of the $e=0.31$ binary. The smaller peaks occur when the binary is at apocenter and initially accretes from the edge of the CBD, and the larger peak is at pericenter. In other words, there are two accretion episodes per orbit. However, this may be a transient feature of the initial relaxation of the fluid; we need a longer term, wider-separation eccentric binary study to establish whether or not this feature is robust. Therefore, we consider this secondary peak a tentative finding.

        The difference between eccentric and quasicircular accretion variability can be explained by a change in the mechanism that fills and depletes their Hill spheres. In the quasicircular case, some works argue that the BHs fill their Hill spheres at a frequency equal to the beat frequency of binary orbit and a `lump' in the CBD, which is an $m=1$ density mode~\cite{shi_three-dimensional_2012, noble_circumbinary_2012}. However, this does not appear to be the mechanism in our simulations as we do not observe a strongly pronounced $m=1$ mode in the CBD's rest-mass density. Note that recent long-term radiation MHD studies of BBH accretion from a CBD show that radiation feedback reduces the appearence of a lump~\cite{tiwari_2025}. Empirically, quasicircular binaries in the strong-field dynamical spacetime regime have been shown to have an accretion frequency of $f \sim 1.4 f_{\rm orb}$~\cite[see, e.g.,][]{paschalidis_minidisk_2021,bright_minidisk,westernacher-schneider_multi-band_2022}, which we confirm in this study. In our simulations, it appears that the eccentricity of the CBD inner edge (or cavity) drives this periodicity, rather than a `lump'. Whenever the BHs approach the minor axis of the elliptical inner disk edge, matter is stripped off the CBD to form accretion streams. This sets the refilling frequency of the minidisks. However, we do not investigate this further, leaving it for a future study. Once filled, the BHs accrete part of the mass in their Hill sphere before it has time to refill, therefore setting the periodicity of accretion at the refilling frequency. In other words, the inflow timescale of the minidisk is shorter than the refilling timescale. At larger BH separations this paradigm can change, as the minidisks will be significantly larger and the inflow timescale can be comparable to, and possibly significantly longer than, the refilling timescale~\cite{bright_minidisk}, thereby suppressing the accretion rate variability significantly.
        
        On the other hand, the eccentric binaries deplete their minidisks at their orbital frequency. We can motivate this behavior by the change in Hill sphere radius on the orbital period. As the binary approaches pericenter, the Hill spheres shrink closer to the ISCO of each BH. Therefore, the BH accretes the matter from its minidisk and any accretion stream plunges in the BH before it has time to circularize into a minidisk. This can be observed in Fig.~\ref{fig:density_inset} where we show rest-mass density contours for our $e=0.31$ simulation just after apocenter (left), about a quarter of an orbit after apocenter (middle), and near pericenter (right), all after the accretion rate has relaxed. We plot the separation-dependent Hill sphere radii with dashed white lines and the ISCO with the solid green lines. We indicate the BH horizons with black disks. At apocenter (largest separation) the binary is in close proximity to the CBD and will refill its Hill sphere, which is at its largest ($r \simeq 6.5 \, M$ for $e=0.17$, $r \simeq 7.2 \, M$ for $e=0.31$). At pericenter, the ratio of the Hill sphere radius ($r \simeq 4.6 \, M$ for $e=0.17$, $r \simeq 3.8 \, M$ for $e=0.31$) to the ISCO radius ($r \simeq 2.475 \, M$) is almost unity. Note that the above mentioned Hill sphere radii correspond to apocenter and pericenter at $t=0$. By the time the accretion rate relaxes (reaches a quasisteady state), the eccentric binaries have inspiraled considerably and the Hill spheres have become even smaller. See the right panel in Fig.~\ref{fig:density_inset} where the ISCO and Hill radius are nearly identical at apocenter ($t/M = 4290.3$). Therefore, matter from the accretion streams plunges into the BHs before it can circularize into minidisks. This both explains the disappearance of minidisks at pericenter (Fig.~\ref{fig:rho_panel}, \ref{fig:density_inset}) and explains the peaks in accretion rate that correspond to pericenter passages (Fig.~\ref{fig:mdot}). 
        
        Additionally, when the accretion rate has settled ($t > 3000 \, M$), the pericenter distance of our eccentric binaries $e=0.17, 0.31$ is $d_{min}\sim11M$ and $14M$, respectively. No persistent minidisks have been found in previous fully general-relativistic MHD simulations of quasicircular binaries at such small separations~\cite{gold_accretion_2014, khan_disks_2018}. A similar Hill-sphere-to-ISCO radius ratio explains the lack of minidisks around retrograde spinning BHs and the significantly smaller minidisks for non-spinning BHs in quasicircular binaries~\cite{paschalidis_minidisk_2021}. This interplay between the Hill sphere radius and ISCO radius occurs on the orbital frequency for eccentric binaries and sets the rest-mass accretion rate periodicity. However, if the eccentric binaries are placed at large enough separation such that, at their pericenter passage, $r_{\rm hill}$ is significantly larger than $r_{\rm ISCO}$, e.g. $r_p \gtrsim 20M$, minidisks may persist throughout the orbit. This requires further study that we will pursue in a future paper.

    \subsection{Jet Launching} \label{subsec:jets}

        \begin{figure*}
            \centering
            \includegraphics[width=\textwidth]{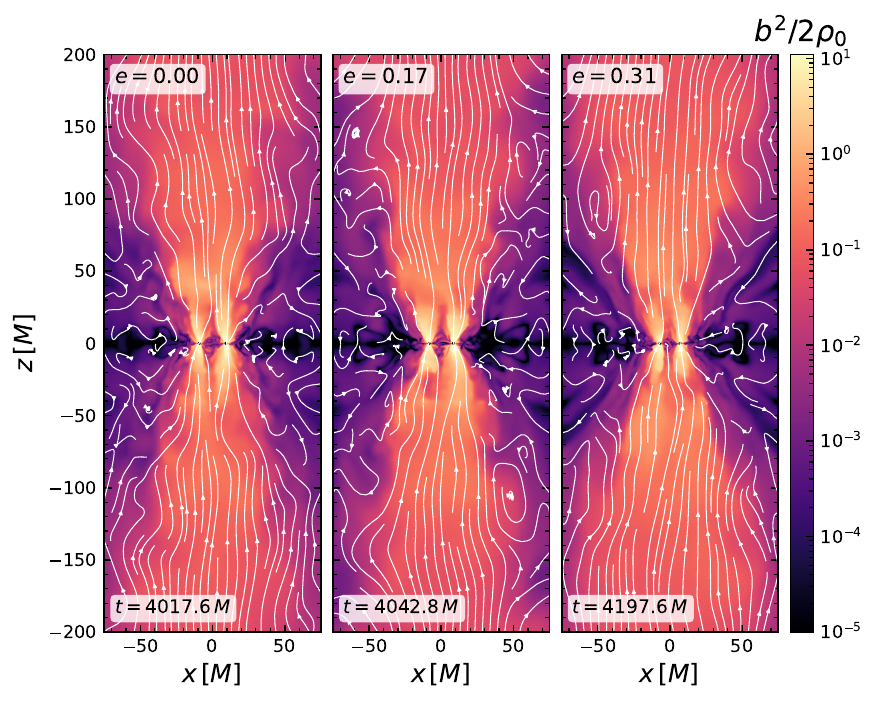}
            \caption{Contours of the fluid magnetization, $\sigma \equiv b^2/2\rho_0$, in log scale. Brighter colors indicate a stronger fluid magnetization. We overlay directed magnetic field lines in white. In each snapshot, the BHs are on the $x$-$z$ plane at $x \sim \pm 10 \, M$ and have strong fluid magnetization directly above and below them with magnetic field lines threading each apparent horizon. From left to right, we plot the three simulations with orbital eccentricities $e={0.00,0.17,0.31}$. All three systems form a dual jet that merges into a single structure for $z \gtrsim 25 \, M$ and are mildly relativistic ($\Gamma \sim 1.15$).}
            \label{fig:jets}
        \end{figure*}

        When ideal magnetized plasma accretes onto a BH, it carries with it a magnetic field that threads the BH event horizon. The interaction of large-scale magnetic fields with the spacetime angular momentum can launch relativistic jets; this is typically referred to as the Blandford-Znajek (BZ) mechanism ~\cite{blandford_electromagnetic_1977}. The energy needed to launch these jets can be provided by the accretion power, the BH rotational kinetic energy, and/or, as in our case, by the orbital kinetic energy of the BBH (as is the case in force-free environments~\cite{Palenzuela:2010nf}).

        \begin{figure*}
            \centering
            \includegraphics[]{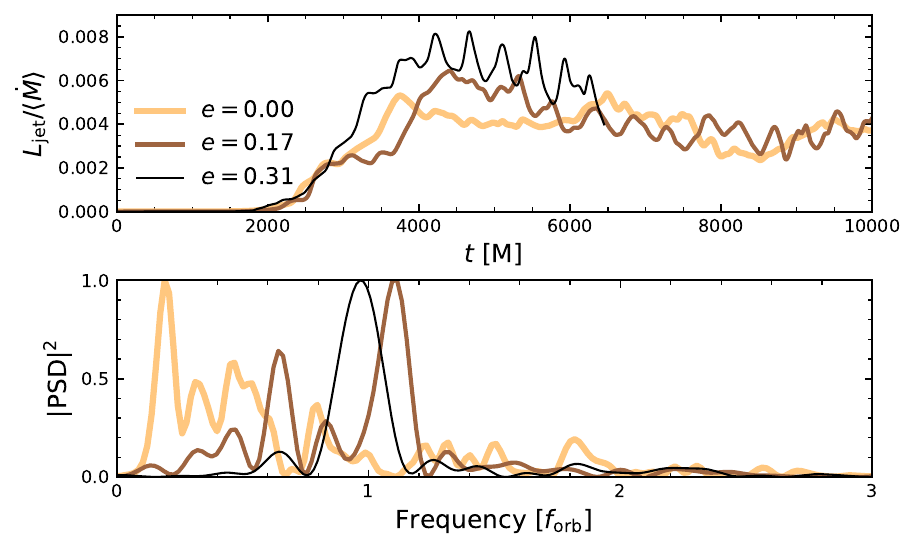}
            \caption{Top: Poynting luminosity normalized to the average accretion rate $L_{\rm jet}/\langle\dot{M}\rangle$ vs time for each binary. The $e=0.00$, $e=0.17$, and $e=0.31$ cases are represented with a thick yellow line, brown line, and thin black line, respectively. The jet luminosity is computed via a surface integral of the the Poynting flux extracted on a sphere with coordinate radius $r=200 \, M$. The time-averaged accretion rate $\langle \dot{M} \rangle$ is measured for $t>3000\, M$. Assuming a $10^7 M_\odot$ BBH accreting at 10$\%$ Eddington, we measure a peak Poynting luminosity of $\sim 10^{43} \, \rm erg \, s^{-1}$. The Poynting luminosity does not reach peak power until $\sim 4000 \, M$, after which it begins to oscillate in the eccentric cases. Bottom: power spectral density of the Fourier transform of the Poynting luminosity shown in the top panel for $t/M > 4000$, with the line style and color representing the same cases. The Fourier transform is performed over the time interval $4000 < t/M < 7000 (6500)$ for the $e=0.17$ ($e=0.31$) binary. The eccentric binaries exhibit Poynting luminosity variability with dominant frequency $f \sim f_{\rm orb}$, similar to their accretion rates. The quasicircular binary has dominant frequency $f \sim 0.2 f_{\rm orb}$.
            }
            \label{fig:poynting}
        \end{figure*}


        Therefore, even though our BHs are nonspinning, we observe jet launching along the $z-$axis of our simulations (the BH orbit is in the $x-y$ plane) consistent with previous studies of nonspinning binary black holes~\cite{farris_binary_2012,Gold_PhysRevD.89.064060,gold_accretion_2014,khan_disks_2018}. In Fig.~\ref{fig:jets}, we plot the fluid magnetization ($\sigma \equiv b^2/2\rho_0$) on a color scale where bright colors are high fluid magnetization and darker colors are low fluid magnetization. On top of the contours, we plot directed streamlines of the magnetic field (using the magnetic field components $B^x$ and $B^z$ measured by a normal observer). We choose moments in our simulation such that the BHs are on the x-axis, and we can plot the $x-z$ plane.

        At $z/M=0$ and $x/M =\pm 10$, we observe two strongly magnetized areas of $\sigma  > 1$ at the base of the jets launched by each BH. Spinning BBHs can have a fluid magnetization that is much greater~\cite{bright_minidisk}. The jet regions, defined by the presence of collimated outflows, extend vertically along the $z-$axis from each BH. At $z/M \gtrsim 25$ the jets from each BH interact and eventually merge to form a single structure. The width of the merged jet increases with vertical distance $z/M$; initially the width is defined by the BH separation and then far away from the BBH ($z/M \sim 200$) the jet has width $\sim 100 \, M$. Within the jet regions the magnetic field is ordered, with vertical field lines extending from the BH horizon out to $z/M > 200$. 

        As we move radially outwards from $x=0$ we enter the bulk of the CBD, where gas density is high and fluid magnetization is roughly $\sigma \sim 10^{-3}$ to $10^{-5}$. In the CBD ($x>25 \, M$), there is a turbulent magnetic field. 
        
        In Fig.~\ref{fig:poynting}, we plot a measure of the jet power efficiency, $L_{\rm jet}/\langle\dot{M}\rangle$, where $L_{\rm jet}$ is the Poynting luminosity measured through a coordinate sphere at $r/M = 200$ and $\langle \dot{M} \rangle$ is the average mass accretion rate computed over $t/M > 3000$. Since the disk only extends out to $r/M \sim 100$, the Poynting flux through this sphere reliably represents the total jet power. In panel A, we plot $L_{\rm jet}/\langle\dot{M}\rangle$ as a function of time. The quasicircular binary is plotted with the thick yellow line, the $e=0.17$ binary with the brown line, and the $e=0.31$ binary with the thin black line. As matter accretes and the initial data relaxes onto the BHs, the jet luminosity increases until it reaches a quasisteady state at $t/M \sim 3700$. Then, the eccentric binaries begin to exhibit variability on their orbital timescale. For a $10^7 M_\odot$ BBH accreting at 10$\%$ Eddington, the peak jet power corresponds to a jet power of $\sim 10^{43} \, \rm erg \, s^{-1}$. In panel B of Fig.~\ref{fig:poynting}, we plot the Fourier transform of these signals. The quasicircular binary ($e=0.00$) has a dominant $0.2 f_{\rm orb}$ frequency in its jet power. While longer evolutions are necessary to confirm this periodicity, this is consistent with previous studies that show the Poynting luminosity variability is not the same as the accretion rate variability~\cite{Combi:2021dks,bright_minidisk}. We note that Ref.~\cite{westernacher-schneider_multi-band_2022} suggested that the jet luminosity variablity would reflect the orbital frequency on the grounds that the  accretion rate periodicity for eccentric binaries matches the orbital period. However, this was not supported by prior works that account for MHD effects in GR. Here, we demonstrate explicitly that the eccentric binaries ($e=0.17, 0.31$) exhibit jet luminosity periodicity $f \sim f_{\rm orb}$. This is the first demonstration that an accreting eccentric BBH has Poynting luminosity variability equal to its accretion rate variability.

    \section{Jet Synchrotron Emission} \label{sec:synchrotron}

    Relativistic electrons within the dual jet can produce synchrotron emission across the EM spectrum~\cite[see, e.g.,][]{cheung2007, walker2018, kim2018}. In this section, we present synthesized spectral energy distributions (SEDs) of the jet synchrotron emission from our simulations (Section \ref{subsec:masseffect}) and their time-dependence (Section \ref{subsec:timesed}).

    Our binaries launch dual jets that merge within a height of $25 \, M$ above the orbital plane (see Fig.~\ref{fig:jets}). Interaction between the two jets can lead to two interesting phenomena, amongst others: the jets can bend at this interaction point, forming a kink instability that results in magnetic reconnection and the efficient acceleration of particles into a nonthermal distribution~\cite{alves_efficient_2018, medina-torrejon_particle_2021}. Furthermore, the merger of the two jets could give rise to a large-scale magnetic reconnection layer where magnetic energy could be efficiently dissipated into a power-law particle energy distribution~\cite{gutierrez_non-thermal_2024}. Particle-in-cell (PIC) simulations have shown that post-reconnection energies are shared, with varying ratios, between the magnetic field, electrons, protons, and positrons, with the electron energy density ranging from $10-100\%$ of the magnetic energy density~\cite{petropoulou_relativistic_2019}. Motivated by these results, in our synchrotron radiation post-processing we adopt a power-law energy distribution and use an electron energy density equivalent to $10\%$ of the magnetic energy density. 
    
    The details of the synchrotron modeling and radiative transfer equation integration are presented in Appendix \ref{sec:synchrotron_appendix}. We summarize our key assumptions here:
    \begin{enumerate}[noitemsep]
        \item As is common, and motivated by the aforementioned studies, we adopt a power-law electron distribution throughout the jet. 
        \item We determine the electron power-law distribution by setting the electron energy density equal to $10\%$ of the magnetic field energy density and the number density of electrons equal to the number density of protons (i.e. charge neutrality).
        \item We begin integrating at $r/M = 50$. This is because we do not perform a general relativistic calculation, therefore our integration of the radiative transfer equation must be in approximately flat spacetime. 
        \item We adopt the ``fast light" approximation; we solve the radiative transfer equation on a slice of constant coordinate time. This approximation is valid when the medium does not change much while light travels through it. In our calculations, it is valid for the optically thin frequencies, which are the main focus here.
        \item We do not treat special relativistic effects. This assumption is consistent with the fact that the fluid bulk velocity in the incipient jets in our simulations is only mildly relativistic ($\Gamma\sim 1.15$), especially at the low heights above the BHs which dominate the synchrotron emission in our calculations.
    
    \end{enumerate}

    Lastly, we report results only from the viewing angle of $\theta = 0$, in other words, looking down the barrel of the jet. We solved the radiation transport equation for other viewing angles and found that the shape and variability of the synchrotron spectrum remain invariant.
    
    \begin{figure*}
            \centering
            \includegraphics[scale=1]{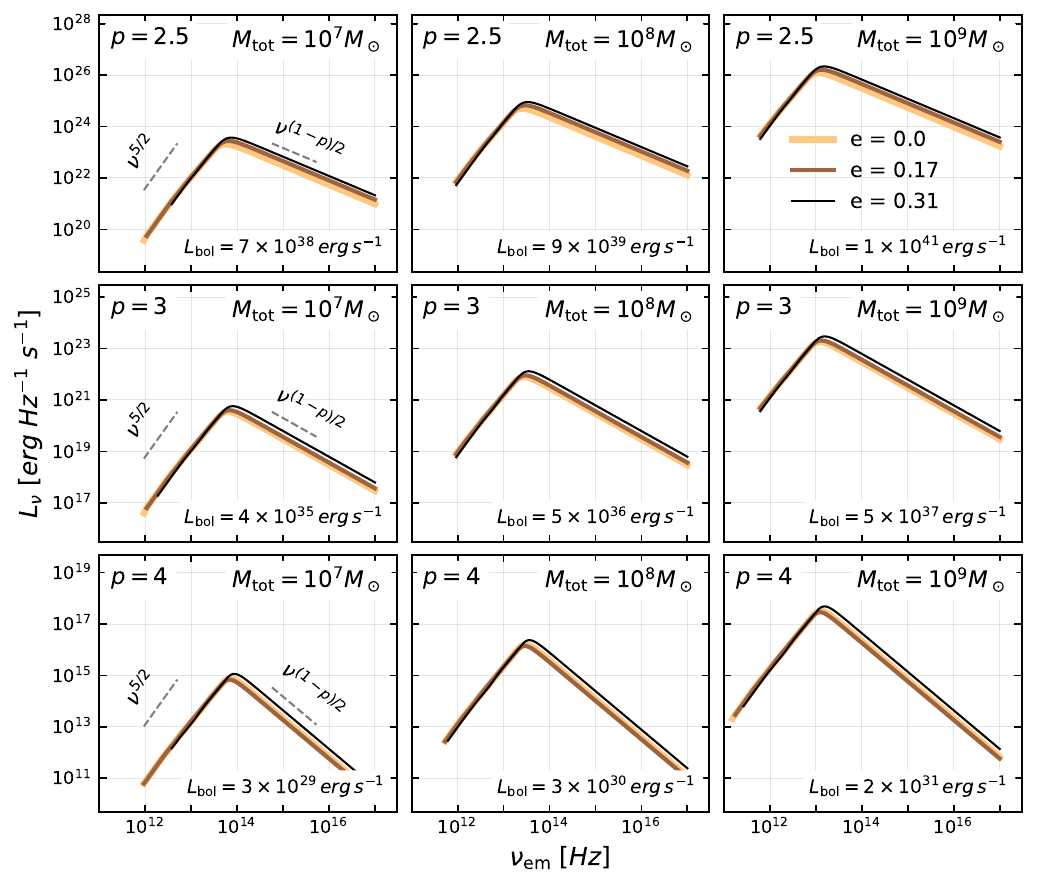}
            \caption{SEDs of synchrotron emission from the SMBBH dual jet. We show the specific luminosity $L_\nu$ as a function of emitted frequency $\nu_{\rm em}$ in the source frame for different parameters. We vary the SMBBH mass in each column; from left to right we select $M_{\rm tot} = 10^7, 10^8, 10^9 \, M_\odot$. The peak luminosity increases with mass ($L_\nu \propto M^{3/2}$) and the peak frequency decreases with mass (also known as the synchrotron self-absorption frequency, $\nu_{\rm ssa}$). We vary the choice of power-law in each row; from top to bottom we show $p = 2.5, 3, 4$, respectively. The peak frequencies are approximately the same in each each column, but the peak luminosities drop as $p$ increases. Lastly, we have also reported the integrated bolometric luminosity of the SEDs in each panel. We note that the peak luminosity would increase by an order of magnitude for spinning BHs, and could go as high as two more orders of magnitude for a full integration down to the BH horizons.}
            \label{fig:sed}
    \end{figure*}

    \subsection{Effect of Binary Mass and Power-law Choice on SED} \label{subsec:masseffect}

    In this subsection, we explore the effect of total BBH mass and power-law of the electron distribution on the calculated synchrotron SED. For the analysis presented here, we set the Eddington factor, $\xi\equiv\langle \dot M\rangle/\dot M_{\rm Edd}$, and radiative efficiency, $\eta$, to $\xi=0.1$ and $\eta=0.1$, respectively. We compute the average accretion rate for $t>3000 \, M$. We note that our adopted accretion rate of $\dot{M} = 0.1 \dot{M}_{\rm Edd}$ to estimate the Synchrotron luminosity may be high enough to cool the accretion disk; however, as we outline below, it is straightforward to scale our results to any accretion rate.

    In Fig.~\ref{fig:sed}, we outline the impact of the power-law and $M$ by performing a radiative transfer calculation through the jet region for a range of masses, $M_{\rm tot} = 10^7, 10^8, 10^9 \, M_\odot$ (columns), and electron distribution power-laws, $p = 2.5, 3, 4$ (rows)~\cite{tsouros2017, gutierrez_non-thermal_2024}. In each panel, we plot the specific luminosity of our BBHs in the source frame as a function of frequency where $e=0.00$ is the thick yellow line, $e=0.17$ is the brown line, and $e=0.31$ is the thin black line. In the figure, it appears that the $e=0.31$ is slightly more luminous that the other cases. However, this is because its average accretion rate after it settles is lower for the same initial disk density and binary mass. Therefore, when we normalize the mass accretion rate to $\dot{M} = 0.1 \dot{M}_{\rm edd}$, its luminosity is higher. This could depend on the initial data chosen, which we will explore in the future.
    
    As we demonstrate in Fig.~\ref{fig:sed}, the specific luminosity increases with increasing BBH mass. It is possible to understand this behavior from analytical considerations of the synchrotron emissivity. In particular, we analytically find that the emission scales with the BBH mass as follows
    \begin{equation}
        L_\nu \sim j_\nu r^{2} \propto  \biggr(\frac{\xi}{\eta}\biggr)^{3/2} M^{3/2},    
    \end{equation}
    For $M=[10^7, 10^9]M_\odot$,
    our radiative transfer calculations show that the SED peak luminosity is $L_\nu \simeq [10^{23}, 10^{26}] \, \rm erg s^{-1}$, in complete agreement with the expected $L_\nu \propto M^{3/2}$ scaling.
    
    Furthermore, as the mass changes in $M=[10^7, 10^9]M_\odot$, the peak frequency falls from $\nu_{\rm ssa} \simeq 5 \times 10^{13}$ Hz to $\nu_{\rm ssa} \simeq 1 \times 10^{13}$ Hz. We use the notation $\nu_{\rm ssa}$ to indicate the synchrotron self-absorption frequency, i.e., the frequency above which the matter becomes optically thin. This behavior can also be analytically derived by finding the frequency at which the optical depth $\tau = 1$. We find the following scaling for $\nu_{\rm ssa}$ with the mass, which matches our radiative transfer calculations
    \begin{equation}
        \nu_{\rm ssa} \propto \biggr(\frac{\xi}{\eta}\biggr)^{\frac{p+6}{2(p+4)}} M^{-\frac{p+2}{2(p+4)}}.
    \end{equation}
    Notice that for $p>2$, the above scaling is very weakly dependent on $p$. 
    
    As we increase the power-law from $p=2.5$ through $p=4$, the peak specific luminosity for $M=10^7 \, M_\odot$ drops from $L_\nu \sim 10^{24} \, \rm erg \, s^{-1} \, Hz^{-1}$ to $L_\nu \sim 10^{15} \rm \, erg \, s^{-1} \, Hz^{-1}$. This is a consequence of how we determine the density constant in the power-law distribution (see equations \ref{eq:electron_distribution}-\ref{eq:min_energy}). As the power-law  increases, the fraction of electrons at the high-energy tail of the distribution falls, which causes the luminosity to drop.

    We integrate the SEDs to determine the bolometric synchrotron luminosity of the BBHs and find that the $M=10^7 \, M_\odot, p=2.5$ BBH has a total synchrotron luminosity of $L_{\rm bol} = 7 \times 10^{38} \, \rm erg \, s^{-1}$. Reference~\cite{bright_minidisk} showed that spinning BBHs can produce a jet Poynting luminosity that is $10$ times greater than what we report in Fig.~\ref{fig:poynting}. Thus, we expect that endowing the BHs with spin would substantially increase the bolometric synchrotron luminosity we find from our synchrotron SEDs.

    \subsection{Detectability of Synchrotron Emission}
    
    The frequencies reported here are within the observing capabilities of NIRCam and MIRI on the James Webb Space Telescope~\cite{greene_2017, miri}, the upcoming Rubin Observatory, the Legacy Survey of Space and Time (LSST), and the Roman Space Telescope~\cite{ivezic_lsst_2019}. However, the bolometric luminosities we report place limits on the distance our predicted synchrotron signals can observed at (see \cite{Manikantan_BBH_eccentric_2024_letter} for details of these calculations). Using the sensitivities of NIRCam, we estimate that NIRCam could observe a $10^7 M_{\odot}$ BBH with $p=2.5$ out to $\sim 0.2 \, \rm Gpc$, which corresponds to $z\sim 0.04$ assuming standard $\Lambda$CDM cosmology. Whereas, it could observe a $10^9 M_{\odot}$ BBH out to $\sim 7 \, \rm Gpc$ ($z\sim 1.1$)~\cite{wright_cosmology_2006, greene_2017}. All these distances would be $\sim 10\times$ greater ($z \sim 7$) if we integrated from $20 \, M$ (where the synchrotron emission is even stronger as we discuss below) instead of $50 \, M$ and took into account the increase in jet power from spinning BHs. Of course, to fully determine detectability of the emission at the frequencies discussed here we would have to consider a comprehensive model that involves emission from the CBD and the minidisks too, which will be the topic of future work. From the pressure in our simulations, we estimated the ion temperature within the CBD to be of $\mathcal{O}({10^{6}}\, \rm K)$. Therefore, the thermal emission frequencies are $\mathcal{O}({10^{16}} \rm \, Hz)$, consistent with previous relativistic works (see e.g.~\cite{Gold_PhysRevD.89.064060}). Such frequencies are $O(100)$ times larger than our reported peak synchrotron frequencies. Therefore, we expect our predicted non-thermal synchrotron spectrum to be more luminous and distinguishable from the thermal spectrum around the peak synchrotron frequencies. However, a full general-relativistic ray-tracing and radiative transfer of these models are necessary to decipher the exact emission from these systems.
    
    \subsection{SED time evolution} \label{subsec:timesed}

    \begin{figure*}
            \centering
            \includegraphics[]{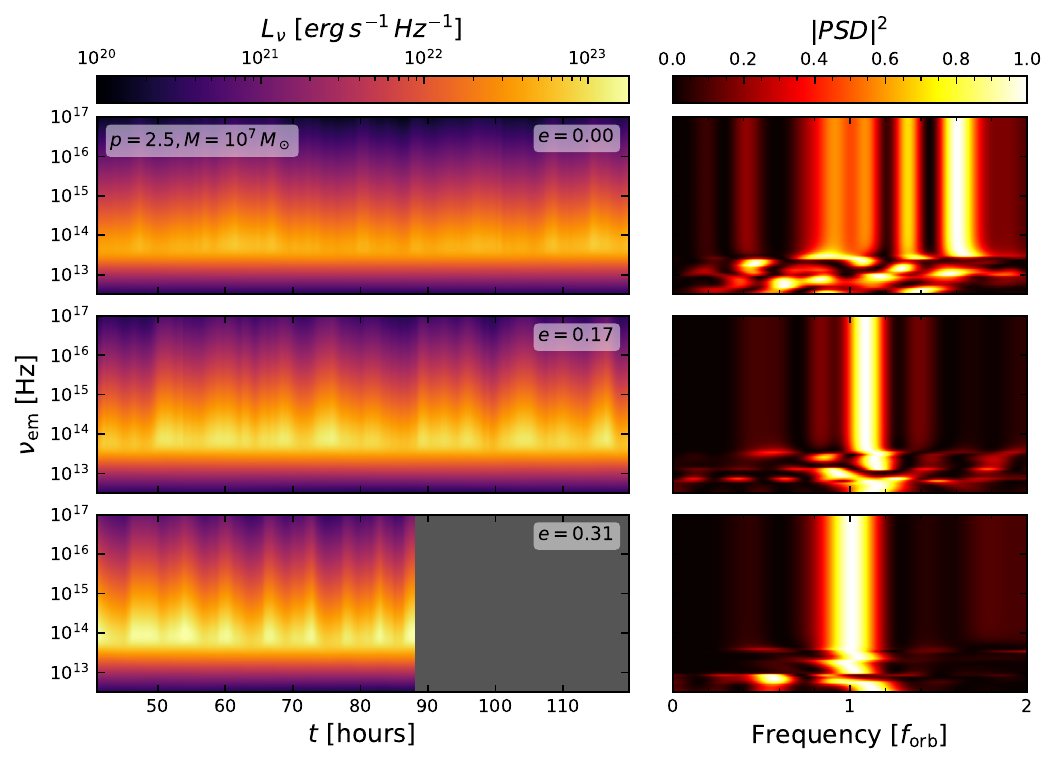}
            \caption{Left: frequency of the synchrotron SED vs time for the $e=0.0$, $e=0.17$, and $e=0.31$ (top to bottom) BBHs. The color map is the specific luminosity with brighter yellow colors indicating a higher luminosity and darker, bluer colors indicating lower luminosity. Right: PSD of the Fourier transform of the frequency-binned SED, where $y$-axis is emission frequency, $x$-axis is the periodicity frequency normalized by the orbital frequency $f_{\rm orb}$, and the color map indicates the strength of the PSD. The PSD shows that only the optically thin region of the synchrotron SED exhibits variability ($\nu \gtrsim 10^{14} \, \rm Hz$. Eccentric binaries have dominant frequency at their orbital frequency $\sim f_{\rm orb}$ and the quasicircular binary of $\sim 1.6 f_{\rm orb}$, which approximately matches their accretion rate periodicity. As in Fig.~\ref{fig:poynting}, we perform the Fourier transform for the time period $t > 4000 \, M$ ($t>54.7 \, \rm hours$), with an upper limit of $7000\, M$ ($96 \, \rm hours$) and $6500 \, M$ ($89 \, \rm hours$) for the $e=0.17$ and $e=0.31$ binaries, respectively. The gray shaded region in the bottom left panel indicates where the $e=0.31$ simulation was stopped because the binary had inspiraled substantially compared to the other cases.}
            \label{fig:inu_fft}
    \end{figure*}

     We performed our synchrotron radiative transfer calculation within the jet with a cadence of $\sim 3.6 \, GM/c^3$ and produced SEDs for each snapshot of constant coordinate time. In Fig.~\ref{fig:inu_fft}, we plot the time-dependent SED (left column) and its frequency-binned Fourier transform (right). From top to bottom, we present the quasicircular ($e=0.00$) and eccentric ($e=0.17$ and $e=0.31$) binaries. The SEDs are all calculated assuming a $10^7 \, M_\odot$ BBH accreting at $10\%$ Eddington, and an electron distribution with power-law $p=2.5$ and electron energy density $10\%$ of the magnetic energy density. 
     
     In the left column of Fig.~\ref{fig:inu_fft}, we describe the time-dependent SEDs with a 2D color map, where the $y$-axis indicates emission frequency and $x$-axis indicates increasing time in days. The brighter yellow color indicates higher specific luminosity ($\sim 10^{24}$) and darker blue/black colors indicate lower specific luminosity ($\sim 10^{21}$). The quasicircular binary shows some variability with time, but without any pronounced structure. The eccentric binaries, however, demonstrate bright `fringe'-like structures where emission peaks. This is most pronounced in the $e=0.31$ binary where we see constantly repeating fringes. Additionally, this variability only appears above the synchrotron self-absorption frequency $\nu_{\rm ssa} \sim 8\times 10^{13} \, \rm Hz$, in the optically thin regime. Below $\nu_{\rm ssa}$, in the optically thick regime, there is no clear variability. Furthermore, the eccentric cases appear to spend more time in a ``low" state, where the synchrotron emission is at a minimum, than in a ``high" state, where the synchrotron emission peaks. Our calculations predict that a smoking-gun synchrotron signature of more eccentric binaries is that they spend longer time in the low state than in the high state. This is consistent with eccentric binaries spending more time at apocenter than pericenter.

     In the right column of Fig.~\ref{fig:inu_fft}, using the measured orbital frequencies (Section \ref{subsubsec:measuringfrequency}), we plot the EM frequency dependent Fourier transforms of the SEDs. The $y$-axis describes frequency of the EM spectrum and the $x$-axis is the frequency of the variability of the specific luminosity. The color map indicates the strength of the PSD, where brighter colors are higher PSD. The eccentric binaries both exhibit variability of $f \sim f_{\rm orb}$ in the optically thin regime (where $\nu \gtrsim \nu_{\rm ssa}$), which is consistent with the variability of their rest-mass accretion rates (Fig.~\ref{fig:mdot}) and their Poynting fluxes (Fig.~\ref{fig:poynting}). The quasicircular binary also exhibits periodicity at $\sim 1.6 \, f_{\rm orb}$, approximately consistent with its rest-mass accretion rate ($\sim 1.4 \, f_{\rm orb}$). There are also secondary harmonics at $\sim 0.9$, $\sim 1.1$, and $\sim 1.35 f_{\rm orb}$.  

     We point that our reported results are not without limitations. The exact values of the luminosities and peak synchrotron frequencies will change with a more careful calculation due to gravitational redshift, among other relativistic effects. However, the qualitative features we report, such as order of magnitude and variability we observe in the optically thin synchrotron regime, are robust and a direct consequence of the varying magnetic field energy density in the jet due to eccentricity. Furthermore, the variability of the eccentric binaries is consistent across choice of power-law, mass, accretion rate, and radiative transfer start point, suggesting that variability on the orbital timescale is a robust feature of non-spinning eccentric binaries with jets and a power-law electron distribution (with electron energy density determined by the magnetic field). By contrast, the SED of the quasicircular binary does not have a consistent dominant frequency in its Fourier transform across the choice of power-law. Furthermore, if we start the radiative transfer at lower heights in the jet, e.g. $z/M \sim 20$, the frequency of variability changes for the quasicircular binary but is unchanged for the eccentric binaries.
     
     Finally, we note that the fast-light approximation, where our rays travel through the medium at an instant of time, is not valid for the optically thick synchrotron spectrum, $\nu \lesssim \nu_{\rm ssa}$. However, the variability we report for the optically thin synchrotron $\nu \gtrsim \nu_{\rm ssa}$ is not affected by this approximation. This is because the jet regions for $z\gtrsim 20M$ remain approximately unchanged over a light-crossing time -- the timescale over which optically thin photons travel through the jet. 

     \subsection{Effect of Integration Starting Point}

     In the previous sections, we reported results where we only integrate from $z>50 \, M$ for self-consistency with our approximations for our radiative transport calculations. We also tested integrations that began at $z=30 \, M$ and $z=20 \, M$. We find that starting at lower heights increases the peak specific luminosity $3\times$ (for $z\geq 30 \, M$) and $10\times$ (for $z\geq 20 \, M$). Moreover, starting at lower $z/M$ shifts the SED towards higher frequencies by a factor of $\sim 2$. However, the shape of the SED and the variability we report in Fig.~\ref{fig:inu_fft} for the eccentric binaries is invariant with the integration start point. On the other hand, the quasicircular binary exhibits variability at the orbital period, but this would need to be confirmed with a fully general-relativistic radiative transfer calculation. We expect curved and dynamical spacetime effects to play progressively more important role for heights $z/M \lesssim 20$, and for this reason we do not perform our radiation transport calculations for $z/M \lesssim 20$. Despite that, we expect that even when one includes these regions, the shape of the SED and the variability we report will remain robust. 

    \begin{figure*}
            \centering
            \includegraphics[width=\textwidth]{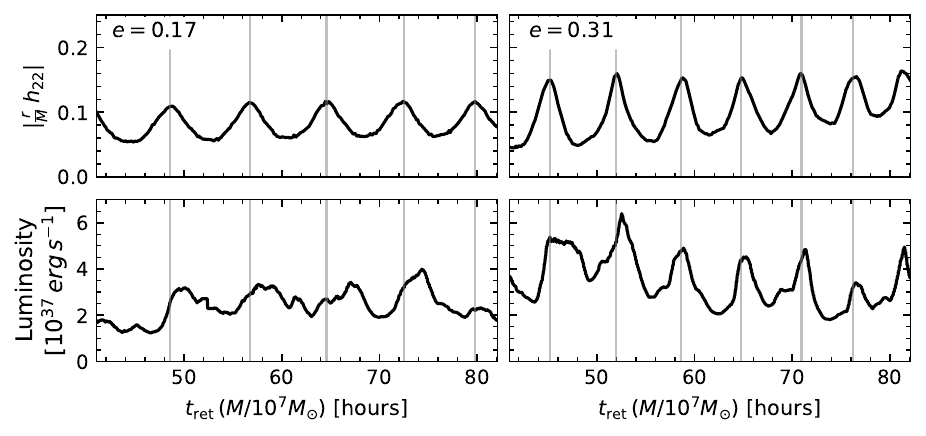}
            \caption{Top row: we plot the amplitude of the $\ell=2, m=2$ mode of the gravitational wave (GW) strain, $\vert h_{22}\vert $, normalized by the distance $r/M$, as a function of retarded time. Bottom row: we plot the synchrotron luminosity integrated for the frequency range $8\times 10^{13} Hz < \nu < 8\times 10^{14} Hz$, as a function of retarded time. This frequency range samples the optically thin synchrotron emission which exhibits robust variability (see Fig \ref{fig:inu_fft}). The time axis is normalized to a $10^7 M_\odot$ binary, as this is the chosen mass for the synchrotron emission plotted. The GW emission exhibits peaks due to binary eccentricity; we denote these peaks with vertical translucent gray lines on both the GW and EM panels. The $e=0.31$ binary shows almost perfect alignment between GW and EM bursts, whereas the $e=0.17$ binary shows about a $100\, M \sim 1.5 (M/10^7M_\odot)\, \rm hours$ delay from GW burst to EM burst.}
            \label{fig:gw_em}
    \end{figure*}

\section{Coincident GW and EM signals} \label{sec:coincident}

    Thus far, we have described novel EM signatures of the accreting binary systems we simulate. However, our simulations also produce self-consistent gravitational waves (GWs) associated with the inspiral of SMBBHs. In this section, we present an analysis of multimessenger signals of the eccentric SMBBH systems we simulated.

    In Fig.~\ref{fig:gw_em}, we demonstrate the simultaneous GW and EM emission from our SMBBHs by plotting the GW strain (top row) and integrated optically thin synchrotron emission (bottom row) for our $e=0.17$ (left column) and $e=0.31$ (right column) binaries. In the top row, we plot the amplitude of the $\ell=2, m=2$ mode of the GW strain, $\vert h_{22}\vert$, scaled with the distance $r$ vs retarded time ($t_{\rm ret}$). In the bottom row, we plot the optically thin synchrotron luminosity integrated for $[\nu \in 8\times 10^{13} {\rm Hz}, 8\times 10^{14} {\rm Hz}]$ vs $t_{\rm ret}$. This frequency range samples the SED near the peak frequency but solely in the optically thin regime, which is variable on the orbital timescale (Fig. \ref{fig:inu_fft}).

    We define the retarded time as $t_{\rm ret}\equiv t-r_0/v$, where $t$ is the coordinate time, $r_0$ is the point where the extraction of the EM or GW signal is performed, and $v$ the speed of propagation of information. For GWs, $r_0$ is the radius of extraction of the GW signal and $v=c$. For the synchrotron emission $r_0$ is the height $z$ at which we start the integration of the radiation transfer equation because, in the fast light approximation, the synchrotron signal propagates to infinity instantaneously, but disturbances in the jet that affect the synchrotron calculation propagate to $z$ in finite time. This time is set by the characteristic velocities in the jet. So, for the synchrotron signal, we take $v=v_{\rm jet}$. Since there is no unique way to define a jet velocity, we define $v_{\rm jet}$ via the time shift necessary to align the phase of Poynting luminosity light curves extracted at different radii in our simulations (e.g. Fig. \ref{fig:poynting} but extracted at $r/M = 100, 130, 150...$ and adjusted in time by $\delta r/v_{\rm jet}$). Doing that, we estimate the jet velocity to be $v_{\rm jet} \sim 0.5$. This jet speed is likely an overestimate of the jet speed for $z \leq 50M$ because we extract the jet Poynting luminosity at $z\gtrsim 100M$, where the jet has accelerated to higher velocities compared to $z \leq 50M$.  However, we emphasize that this is not the only way to measure jet velocity. If, instead, we choose to average the fluid velocity in highly magnetized regions at the base of the jet, we find that the jet velocity is $v_{\rm jet} \sim 0.1-0.2$. Regardless, the calculation of the retarded time for both the EM and GW signals is only approximate and is only meant to help gain some understanding on the coincidence of GW and EM signals we expect from these sources.

    In the $e=0.17$ binary (left column in Fig.~\ref{fig:gw_em}), we find that the GW bursts precede the EM bursts by about $t/M \sim 100$ or $t \sim 1.4 (M/10^7 M_\odot) \rm \, hours$. If we instead set $v_{\rm jet}\sim 0.2$, we find that the EM and GW peaks line up almost exactly. In the $e=0.31$ binary (right column in Fig.~\ref{fig:gw_em}), using $v_{\rm jet}=0.5$ as determined by the Poynting luminosity measurements, we find that the GW and EM bursts happen almost simultaneously; in certain cases the GW bursts marginally precede the EM bursts (see the second, third, and fourth peaks). The key takeaway from this plot is that the time period between successive bursts is the same for both the GW and EM synchrotron emission. This is the defining feature of binaries that have non-negligible eccentricity. 

    A fully general-relativistic radiative transfer calculation free of the fast-light approximation down to the BBH horizons is necessary to draw robust conclusions on the simultaneity of, or lag between, the time of arrival of the EM and GW bursts. However, our main conclusion here is robust: binary eccentricity should manifest itself such that the time period between successive bursts is the same for both the GW and EM synchrotron emission. 

\section{Summary and dicussion} \label{sec:summary}

    In this work, we reported results from the first systematic investigation of magnetohydrodynamic accretion onto eccentric BBHs in full 3+1 general relativity. We explored eccentricities of $e = 0.00, 0.17$, and $0.31$ at a binary major axis of $a/M=20$. We simulated the $e = 0.00$ and $0.17$ binaries out to $t/M \sim 10,000$ (19 and 18 orbits, respectively) and the $e=0.31$ binary to $t/M \sim 6500$ (13 orbits). 
    
    We embedded the binaries in the central cavity of an initially geometrically thick torus which we seeded with a weak poloidal magnetic field to render it unstable to the magnetorotational instability. This instability leads to turbulence in the disk and acts as an effective viscosity that drives accretion onto the binary. Tidal streams from the inner-edge of the CBD fill the Hill spheres of the BHs to form intermittent minidisks which quickly accrete in the eccentric cases, but are more persistent in the quasicircular case. This process periodically repeats. Accretion of the magnetized matter also enables the BHs to launch collimated outflows -- jets -- despite the individual BHs not spinning. Our key findings are listed below:

    \begin{enumerate}[noitemsep]
        \item The accretion rate onto eccentric binaries varies with time and exhibits periodicity with frequency $f \sim f_{\rm orb}$, unlike quasicircular binaries whose accretion rate variability exhibits peak periodicity at $f \sim 1.4 f_{\rm orb}$. 
        \item Quasicircular binaries at separation $d=20 \, M$ have a persistent minidisk structure throughout their orbit. However, eccentric binary minidisks are quickly depleted at pericenter for the relativistic separations we study here. Consequently, quasicircular binaries have persistent nodes of low density in their cavity that are an order of magnitude less dense than the cavity of eccentric binaries. 
        \item Eccentric binaries launch jets with a Poynting luminosity that exhibits periodicity with frequency $f \sim f_{\rm orb}$, while that of quasicircular binaries exhibits periodicity with $f \sim 0.2 f_{\rm orb}$. However, the latter variability is not as pronounced and requires longer simulations to confirm.
        \item Optically thin synchrotron emission from the jet exhibits variability $f\sim f_{\rm orb}$ for eccentric binaries. This variability is agnostic to the choice of electron distribution power-law as long as we determine the electron energy density to be a fraction of the magnetic energy density.
        \item The delay between consecutive gravitational wave bursts and the delay between consecutive optically thin synchrotron emission bursts is the same. This is a smoking gun signature of SMBBHs with non-negligible eccentricity. 
    \end{enumerate}

    Our suite of simulations and the results derived from them are most applicable to sub-Eddington SMBBHs as we do not include radiation feedback. The accretion rate periodicity we find for quasicircular binaries, $f \sim 1.4 f_{\rm orb}$, is consistent with previous studies. The periodicity of the eccentric binaries, $f \sim f_{\rm orb}$, however, is a new result that we first reported in~\cite{Manikantan_BBH_eccentric_2024_letter} for $e=0.31$, and have confirmed here with an even lower value of eccentricity $e=0.17$. This periodicity is driven by pericenter passages, where the Hill sphere of each BH shrinks and approaches the size of the innermost circular orbit of each BH. This drives accretion onto each horizon. Additionally, the peaks in accretion rate correspond directly to pericenter passages, which further supports this idea. It is reasonable to expect that there is a threshold value for the eccentricity above which binaries exhibit periodicity on the orbital frequency and below which binaries exhibit periodicity at $1.4\times$ the orbital frequency. It is also possible that the periodicity decreases continuously between $1.4 \times f_{\rm orb}$ and $f_{\rm orb}$ as $e$ increases from 0 to a threshold value. However, we do not establish at which value of $e$ this occurs in this work. Furthermore, we find that the Poynting luminosity of eccentric BBHs oscillates at the orbital frequency, consistent with its accretion rate periodicity. This is in contrast to quasicircular binaries where we find a periodicity of $\sim 0.2 f_{\rm orb}$ and other studies find varying periodicities~\cite{Combi:2021dks, bright_minidisk}.

     Additionally, we perform an approximate radiative transfer calculation of the jet synchrotron emission assuming the existence of electrons with a power-law energy distribution. We find that the optically thin synchrotron emission has variability consistent with the binary orbital frequency. This variability is insensitive to the choice of electron distribution power-law. However, as we found in~\cite{Manikantan_BBH_eccentric_2024_letter}, we also find that the variability is sensitive to how we set the energy in the power-law distribution. If we do not assume that the electron and magnetic energy density are linked, then variability is not as clear. Observations of eccentric binaries that confirm or deny jet variability for non-thermal electrons could give us further insight into how electron distributions adapt to changing magnetic field energies. Nevertheless, the equipartition between the electron energy and magnetic field energy is well-established~\cite{scott_low-frequency_1977, chevalier_synchrotron_1998, panaitescu_properties_2002}. In addition to the periodicity that matches the orbital frequency, a smoking-gun signature of the emission from eccentric binaries is that they spend more time in a low state, where the synchrotron emission is at a minimum, than in a high state, during which the synchrotron emission peaks. A second smoking-gun feature is the coincident EM and GW bursts from the eccentric binaries. While a more careful calculation is needed to establish these bursts as simultaneous, a unique feature of eccentric binaries is that the time between their successive synchrotron bursts is the same as the time between corresponding GW bursts.
    
    Additionally, we showed how the total luminosity and peak frequency of our synchrotron SEDs vary with choice of total binary mass and the power-law exponent. For an exponent of $p=2.5$ and $M_{\rm tot} = 10^9 \, M_{\odot}$, the SMBBH has a bolometric luminosity of $L_{\rm bol} = 1 \times 10^{41} \, \rm erg s^{-1}$, suggesting that an instrument such as NIRCam would observe such an object out to $\sim 7 \, \rm Gpc$. Spinning BHs could increase the total luminosity by an order of magnitude~\cite{bright_minidisk}. Furthermore, calculations that started closer to the horizon ($z \sim 20 \, M$ instead of $z \sim 50 \, M$) can increase the peak luminosity by a further order of magnitude, making these systems potentially detectable at large cosmological redshifts.
    
    A fully general-relativistic ray-tracing and radiative transfer calculation is necessary to decipher the exact emission from the jet region and to understand emission from processes in the circumbinary disk, the inner cavity, and the minidisks. The latter is expected to be responsible for the bulk of the X-ray/UV emission as well as Doppler shifted emission lines~\cite{sesana_multimessenger_2012, charisi_multi-messenger_2022, bogdanovic_bhb_review}.
    
    Our simulations are an essential first step towards understanding the behavior of eccentric BBHs in the GW-driven regime. We have revealed novel features in their accretion rate and jet emission. However, these simulations have their limitations; for example, we have only considered a non-spinning binary embedded in an aligned and thick accretion disk. An exploration of the full parameter space including configurations with spins for the BHs, unequal masses, tilting the accretion disk with respect to the binary's angular momentum, and exploring different accretion regimes with a thinner accretion disk, are necessary to achieve a complete understanding of the dynamics of and emission from accreting SMBBHs. Finally, radiation feedback effects have not been considered and can play a substantial role in cooling and driving winds from the CBD and minidisks when near the (super-) Eddington regime. In future work we will investigate all these parameters.

\begin{acknowledgments}
    This work was in part supported by NASA grant 80NSSC24K0771 and NSF grant PHY-2145421 to the University of Arizona. We thank Collin Christy and Thomas Baumgarte for useful discussions. This research is part of the Frontera computing project at the Texas Advanced Computing Center. This research was in part supported by ACCESS under allocation award No. PHY190020. Frontera is made possible by National Science Foundation award OAC-1818253. ACCESS is supported by National Science Foundation grants 2138259, 2138286, 2138307, 2137603, and 2138296~\cite{boerner_access_2023}. 
\end{acknowledgments}

\appendix
\section{Measuring Orbital Eccentricity} \label{sec:eccentricity_measurements}

    \begin{figure*}[ht]
        \centering
        \includegraphics[scale=0.8]{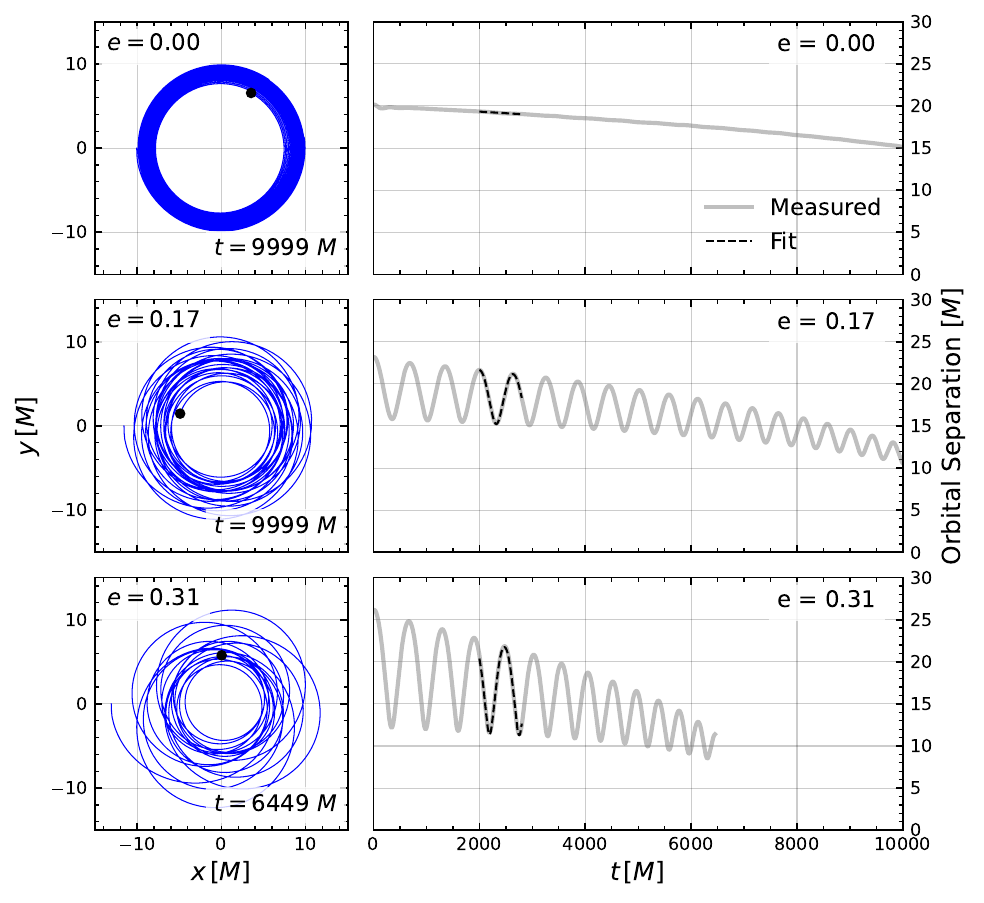}
        \caption{The orbits of our non-spinning, binary black holes with measured eccentricities of $0.00, 0.17,$ and $0.31$. In the left column, we plot the $x-y$ positions of one apparent horizon. The $z-$position is constant as there is no vertical precession in the orbits of non-spinning BHs. In the right column, we plot the time-dependent orbital separation of the BHs with the thick gray lines and the best fit of Equation \eqref{eq:separation_fit} with the dashed black line. The orbital eccentricity fit is performed for the time interval $2000 < t/M < 3500$. The upper end of this interval is close to the time at which the accretion rate onto the binaries settles. We indicate the measured eccentricities in the top left and right of the panels.}
        \label{fig:orbits}
    \end{figure*}
    
        The definition of eccentricity in the fully general relativistic two-body problem is not straightforward (see, e.g.,~\cite{Shaikh:2023ypz, boschini_orbital_2024}).
        In particular, there exists no unique definition for eccentricity, and some definitions are not gauge independent. Here, we use a simple approach based on the quadrupole approximation to approximately fit for the eccentricity of our orbits. Our approach is not gauge invariant but it is easy to implement and yields reasonable results. 
        
        We start with the solution of the Newtonian 2-body problem, where the orbital separation evolves in time as 
        \begin{equation} \label{eq:separation}
            d(t) =\frac{a(t)(1-e(t)^2)}{1+e(t)\cos{(\omega t + \varphi})}.
        \end{equation}
        Here $a(t)$ is the time-dependent semi-major axis due to the inspiral of the black holes, $e$ is the (generally) time-dependent eccentricity of the orbit, $\omega$ is the angular frequency of the orbit, and $\varphi$ is a phase offset. Since we are in a regime where energy dissipation occurs through gravitational waves, the average power over one period of an elliptical orbit is given by~\cite{peters_matthews} 
        
        \begin{equation} \label{eq:Lgw}
            L_{GW} = \frac{32}{5} \frac{G^4}{c^5} \frac{M^3\mu^2}{a^5} f(e)
        \end{equation}
        where $M = m_1 + m_2$, $\mu = m_1m_2/(m_1+m_2)$, and $f(e)$ is 
        \begin{equation}
            f(e) = \frac{1+73/24e^2 + 37/96e^4}{(1-e^2)^{5/7}}.
        \end{equation}

        Finally, by equating the gravitational wave power to the time derivative of the orbital energy, we can solve for the time derivative of the semi-major axis
        \begin{equation*}
            \frac{dE}{dt} = \frac{GM \mu}{2a^2} \dot{a} = -\frac{32}{5} \frac{G^4}{c^5} \frac{M^3\mu^2}{a^5} f(e)
        \end{equation*}
        \begin{equation}
            \Rightarrow \frac{da}{dt} = \frac{-64G^3}{5c^5} \frac{M^2\mu}{a^3}f(e)
        \end{equation}

        Integrating the above equation (assuming $e$ is constant) ones finds
        \begin{equation}\label{gw_a}
            a(t) = a_0 \bigl(1-\frac{4t}{\tau}\bigl)^{1/4},
        \end{equation}
        where $a_0$ is the initial semi-major axis of the orbit and $\tau$ is the coalescence timescale of the binary (which effectively contains $f(e)$ when $e$ is constant). We can substitute Equation \eqref{gw_a} into Equation \eqref{eq:separation} to obtain
        \begin{equation}\label{eq:dvst}
            d(t) = a_0 \bigl(1-\frac{4t}{\tau})^{1/4} \frac{(1-e^2)}{1+e\cos{(\omega t + \varphi})}.
        \end{equation}

        Finally, due to non-negligible eccentricity of the BBH orbits, we include the true anomaly $\theta$ in the $\cos$ in the denominator of Equation~\eqref{eq:dvst}, but we  expand it in eccentricity to $O(e^3)$~\cite{smart}
        \begin{equation}\label{true_anomaly}
            \theta = \psi + \Bigl(2e-\frac{1}{4}e^3\Bigl)\sin{\psi} + \frac{5}{4} e^2 \sin{2\psi} + \frac{13}{12} e^3\sin{3\psi}.
        \end{equation}
        Here $\psi = \omega t + \varphi$ is the mean anomaly. The mean anomaly and the true anomaly are equal for circular orbits, but differ for eccentric orbits. Therefore, the final equation we use to find the eccentricity in our simulations is

        \begin{equation}
        \label{eq:separation_fit}
            d(t) =a_0 \bigl(1-\frac{4t}{\tau})^{1/4}\frac{(1-e^2)}{1+e\cos{(\theta})},
        \end{equation}
        with $\theta$ given by Eq.~\eqref{true_anomaly}. We are left with five parameters which we fit for using the coordinate separation of the binaries in our simulations: $\tau$, the coalescence timescale, $a_0$, the initial semi-major axis, $\omega$, the orbital frequency, $\varphi$ the phase, and the eccentricity, $e$.

        In Figure~\ref{fig:orbits}, we plot our three binaries with varying eccentricity, $e=[0.00, 0.17, 0.31]$. In the left column, we plot the $x-y$ orbital tracks of the BBH from initial conditions through to $t/M\sim 10000$. We plot the orbital track for just one of our BHs because the BHs are identical and have identical orbital tracks. In the right column, we plot the orbital separation of the BBH as a function of time. The solid gray line represents the orbital separation in the simulation and the dashed black line describes our best fit as determined by the method outlined above. 
        
        The top row in Fig.~\ref{fig:orbits} describes the quasicircular orbit, where $e\sim 0.0$. As expected, the orbital separation is approximately constant over an orbit ($\sim 500 \, M$) and decays slowly over the GW timescale. The middle row describes the $e=0.17$ binary. The initial orbital separation varies from $d=23M$ to $17M$ and the semi-major axis decays from $a=20\, M$ to $\sim 15 \, M$. The bottom row describes our highest eccentricity binary, where $e=0.31$. The initial orbital separation varies from $d=26M$ to $14M$ across an orbital timescale and the semi-major axis decays from $a=20\, M$ to $\sim 10 \, M$. As expected the eccentric binaries decay faster (see~\cite{peters_matthews} for an analytic description). If we include the leading order correction to the eccentricity evolution from~\cite{peters_matthews}, our results are unchanged for the $e=0.00, 0.17$ cases. The $e=0.31$ simulation, however, has a best fit value of that is $\lesssim 1\%$ larger than the value measured without the leading order correction in the evolution of the eccentricity.

        We also implement the \texttt{gw\_eccentricity} package introduced in~\cite{Shaikh:2023ypz} to measure the eccentricity of our binaries with their $\ell,m=(2,2)$ gravitational wave mode. We measure our $e=0.17$ binary to have an eccentricity of $e\sim0.15$ at $t/M \sim 1500$ and $e\sim0.13$ at $t/M\sim2000$. We measure our $e=0.31$ binary to have an eccentricity of $e\sim0.35$ at $t/M\sim 1500$ and $e\sim0.29$ at $t/M \sim 2000$. This is generally in agreement with what we measure with the orbital separation method we outlined. But, we find that the GW method has too much variance with respect to when the eccentricity measurement is performed. This is likely due to the fact that our waveforms have some noise. Therefore, we use our orbital separation eccentricity measurements throughout to label our cases. 

\section{Synchrotron Emission} \label{sec:synchrotron_appendix}

    We perform approximate synchrotron radiative transfer assuming flat spacetime and ignoring special relativistic effects of the bulk fluid. We start our integrations at large heights $z\geq 20-50M$ above the orbital plane, where the spacetime metric is approximately described by the Minkowski metric, to justify this approximation. The plasma velocities in the jet region are only mildly relativistic $v\sim 0.2-0.3c$. Therefore, we expect only $O(10\%)$ corrections to our results with a more careful computation. In an upcoming work, this computation will integrate the covariant radiative transfer equation all the way down to the horizons. 
    
\subsection{Radiative Transfer Equation} \label{subsec:rte}
    
    We use the definitions for synchrotron emissivity $j_\nu$ and absorption coefficient $\alpha_\nu$ of~\cite{rybicki_lightman},
    \begin{equation}
        \begin{aligned}
            j_\nu \,= \, &\frac{\sqrt{3}q^2 C B \sin{\alpha}}{m_e c^2 (p+1)} \biggr(\frac{m_e c \nu}{3qB\sin{\alpha}}\biggr)^{-(p-1)/2} \\
            &\Gamma\biggr(\frac{3p+19}{12}\biggr) \Gamma\biggr(\frac{3p-1}{12}\biggr),
        \end{aligned}
    \end{equation}
    \begin{equation}
        \begin{aligned}
            \alpha_\nu \,=\, &\frac{\sqrt{3}q}{8 \pi m_e}\biggr(\frac{3q}{2 \pi m_e^3 c^5}\biggr)^{p/2} C (B\sin{\alpha})^{(p+2)/2} \\
            &\Gamma\biggr(\frac{3p + 2}{12}\biggr)\Gamma\biggr(\frac{3p + 22}{12}\biggr)\nu^{-(p+4)/2},
        \end{aligned}
    \end{equation}
    where $q$ is the electron charge, $m_e$ is the electron mass, $c$ is the speed of light, $B$ is the magnetic field strength, $\nu$ is the frequency of light, $\Gamma$ is the special function, $C$ and $p$ are used to describe the electron energy distribution, and  $\alpha$ is the pitch angle. In the calculations we perform, we integrate $\alpha$ out of the equation by assuming that the electrons follow an isotropic pitch angle distribution between $0<\alpha<\pi/2$. Furthermore, we assume that the electrons follow a power-law energy distribution:
    \begin{equation} \label{eq:electron_distribution}
        N(E) \, dE = C E^{-p} \, dE,
    \end{equation}
    where $p$ is the chosen power-law of the electron distribution and $C$ is a density constant. We can determine $C$ by specifying the  electron number density and  energy density using our simulations. First, we assume charge neutrality. This implies
    \begin{equation}\label{conC}
        \begin{aligned}
            \int^{\infty}_{E_1} &N(E) dE = \int^{\infty}_{E_1} CE^{-p}dE = \rho/m_p \\
            &\Rightarrow C = (p-1)E_1^{(p-1)} \rho/m_p,
        \end{aligned}
    \end{equation}
    where $\rho$ is the rest-mass density measured in our simulations and $m_p$ is the mass of the proton. We assumed that the matter is made of fully ionized Hydrogen. We have introduced an additional parameter, $E_1$, which is the minimum energy of an electron in this power-law distribution.  Here, we choose $E_1$ by assuming some level of equipartition with the local magnetic field. In particular, we assume that the energy density of the local electron distribution is fixed to 10\% of the local magnetic energy density. This estimated fraction is consistent with particle-in-cell simulations of magnetic reconnection (see, e.g.,~\cite{petropoulou_relativistic_2019}). The equipartition assumption is also motivated by earlier studies~\cite{scott_low-frequency_1977, chevalier_synchrotron_1998, panaitescu_properties_2002}, and has been used in analysis of astronomical synchrotron sources, e.g., signals from tidal disruption events in~\cite{christy_peculiar_2024}. Thus, we set
    \begin{equation}
        \epsilon_e = \zeta \epsilon_B,
    \end{equation}
    where the electron energy density is
    \begin{equation}    
        \begin{aligned}
        \epsilon_e = \int^{\infty}_{E_1} E\, N(E&)\, dE = \int^{\infty}_{E_1} CE^{(-p+1)} dE \\
        \Rightarrow \epsilon_e &= \frac{C}{p-2}E_1^{-(p-2)},
        \end{aligned}
    \end{equation}
    and the magnetic field energy density is
    \begin{equation}
        \epsilon_B = B^2/8\pi,
    \end{equation}
    with $B$ the magnetic field strength measured by an observer comoving with the plasma. Solving for $E_1$
    yields
    \begin{equation} \label{eq:min_energy}
        E_1 = \frac{p-2}{p-1} \frac{\zeta B^2 m_p}{8 \pi \rho}.
    \end{equation}
    Using the last equation and Eq.~\eqref{conC} we determine for the constant $C$. We assume $\zeta=0.1$ for the calculations in this work. However, we stress that this choice does not affect the variability of our synchrotron emission.

    Finally, we integrate the time-indepedent radiative transfer equation~\cite{rybicki_lightman} 
    \begin{equation} \label{eq:radtran}
        \frac {dI_\nu}{d\ell} = (j_\nu - \alpha_\nu I_\nu),
    \end{equation}
    along a line of sight, using a simple forward Euler scheme with resolution $\Delta\ell = 2\, M$. We have checked for convergence and that our reported results are invariant with resolution.

    \subsection{Implementation} \label{subsec:synchrotron_implementation}

    We consider the special case where the observer is directly along the line of sight of the jet (i.e. $\theta=0$). In our simulation, this means the observer is at $x=y=0$ at some vertical distance above the BBH. We integrate along one line of sight and consider the emitting region to be $x,y \pm 250 \, M$ such that the intrinsic luminosity is the intensity multiplied by the area of the emitting region. To scale our simulation to the supermassive black hole systems of interest, we set the average rest-mass accretion rate to 10\% the Eddington accretion rate, that is,

    \begin{equation}
        \dot{M}_{\rm edd} = \frac{L_{\rm edd}}{\eta c^2} = \frac{4 \pi G M m_p}{\eta\sigma_Tc}
    \end{equation}
    where $\eta$ is the radiative efficiency of the accretion, $M$ is the gravitational mass of the binary, $m_p$ is the proton mass, and $\sigma_T$ is the Thompson cross section.

    Our accretion disk is geometrically thick, making it an appropriate model for sub-Eddington accretion. Therefore, we choose an $\langle\dot{M}\rangle = \xi \dot{M}_{\rm edd}$, with $\xi=0.1$. While $\xi=0.1$ is still low enough to completely neglect radiation feedback, we adopt it as an upper bound to the validity of our model. One can straightforwardly use a different value since our calculations scale with $\xi$. We also assume a radiative efficiency of $\eta=0.1$ for the calculations reported in this work\footnote{Note that while the total luminosity output from our calculations depends on $\eta$, our reported periodicities do not.}. Therefore, our desired accretion rate is
    \begin{equation}\label{eq:Mdot_desired}
        \dot{M}_{\rm desired} = 7 \times 10^{-10} \left(\frac{\xi}{0.1}\right)\left(\frac{\eta}{0.1}\right)^{-1}\left(\frac{M}{10^7M_\odot}\right) M_\odot / s.
    \end{equation}

    The next step is to convert quantities from the geometrized units of our code to cgs units that correspond to the desired accretion rate. If, for example, the measured time averaged rest-mass accretion rate is $\langle\dot{M}\rangle = 1$, we convert the simulation accretion rate from geometrized units with $M=1$ to cgs units by multiplying by $c^3/G$, which gives us 
    \begin{equation*}
        \dot{M}_{\rm cgs} = 1 \times c^3/G \simeq 2 \times 10^5 M_\odot/s. 
    \end{equation*}

    Finally, we scale the density and magnetic field strength such that the simulations have the specified time averaged accretion rate of Eq.~\eqref{eq:Mdot_desired} for the time period $t/M > 3000$, i.e., after the accretion rate settles. We follow the above procedure to find this scaling factor for each of our simulations, and use it to scale the densities and magnetic fields. We find that the maximum density at $t=0$ for the quasicircular binary is given by 
\begin{widetext}
    
    \begin{equation}
        \rho_{0, \rm max} = 1.6 \times 10^{-11} \biggr(\frac{\xi}{0.1 }\biggr) 
        \biggr(\frac{M}{10^7 M_{\odot}}\biggr)^{-1} 
        \biggr(\frac{\eta}{0.1}\biggr)^{-1} \rm \, \frac{g}{cm^{3}},
    \end{equation}
    while for the $e=0.17$ and $e=0.31$ cases it is given by

    \begin{equation}
        \rho_{0, \rm max} = 1.8 \times 10^{-11} \biggr(\frac{\xi}{0.1 }\biggr) 
        \biggr(\frac{M}{10^7 M_{\odot}}\biggr)^{-1} 
        \biggr(\frac{\eta}{0.1}\biggr)^{-1} \rm \, \frac{g}{cm^{3}},
    \end{equation}
    and
    \begin{equation}
        \rho_{0, \rm max} = 2.6 \times 10^{-11} \biggr(\frac{\xi}{0.1 }\biggr) 
        \biggr(\frac{M}{10^7 M_{\odot}}\biggr)^{-1} 
        \biggr(\frac{\eta}{0.1}\biggr)^{-1} \rm \,\frac{g}{cm^{3}},
    \end{equation}
respectively. The amplitude squared of the magnetic scales like the density, so we can use the previous expression to scale the magnetic fields for each of our simulations, and use them as input to our synchrotron radiation transport. 
\end{widetext}

\bibliography{sample631}

\end{document}